%% file: neurips_2026.tex
\PassOptionsToPackage{numbers}{natbib}
\documentclass{article}

\usepackage[preprint]{neurips_2026}

\usepackage[utf8]{inputenc} 
\usepackage[T1]{fontenc}    
\usepackage{hyperref}       
\usepackage{url}            
\usepackage{booktabs}       
\usepackage{amsfonts}       
\usepackage{nicefrac}       
\usepackage{microtype}      
\usepackage{xcolor}         
\usepackage{graphicx}

\usepackage{multirow}
\usepackage{booktabs}

\usepackage{makecell}
\usepackage{amsmath}

\usepackage{array, tabularx}

\usepackage[most]{tcolorbox}

\usepackage{capt-of}
\usepackage{enumitem}
\usepackage{pifont}

\title{\textsc{PDAgent-Bench}: Characterizing, Grounding, and Architecting LLM/VLM Agents for VLSI Physical Design}

\author{
  Qiufeng Li$^{1, *}$, Rongqian Chen$^{1}$, Quan Cheng$^{2}$, Chengxuan Wang$^{3}$, 
  Sizhe Tang$^{1}$ \\ 
  Chia-Tung Ho$^{4}$, David Z. Pan$^{5}$,  
  Tian Lan$^{1}$, Weidong Cao$^{1,*}$ \\
  $^{1}$ George Washington University \quad $^{2}$ Brown University 
  $^{3}$ University of California, Los Angeles  \\
  \quad 
  $^{4}$ NVIDIA \quad
  $^{5}$ The University of Texas at Austin  \\
  {\texttt{\{qiufeng.li, weidong.cao\}@gwu.edu}} \\
}

\begin{document}

\maketitle

\begin{abstract}

Large Language Models (LLMs) and vision-language
models (VLMs) have shown remarkable success in the front-end design of Very Large-Scale Integrated Circuits (VLSI), yet their capabilities for VLSI physical design remain significantly underexplored. 
One primary cause is the lack of standardized benchmarks for evaluating agentic physical design workflows that require high-dimensional, multi-stage optimization under strict design constraints, coordinated interaction with diverse Electronic Design Automation (EDA) tools, and iterative refinement.
This work introduces \textsc{\textbf{PDAgent-Bench}}, a comprehensive and multi-dimensional \underline{\textbf{bench}}mark for evaluating LLM/VLM-based \underline{\textbf{agent}}s across the \underline{\textbf{p}}hysical \underline{\textbf{d}}esign stack.
\textsc{PDAgent-Bench} integrates both task-level assessment and workflow-level execution.
The benchmark suite contains 353 curated problems that combine conceptual questions with real-world industrial artifacts, with expert-validated references and executable solutions.
\textbf{These tasks cover five key capability dimensions}: foundational knowledge, report comprehension, root-cause analysis, script generation, and full-flow implementation (including \textbf{macro placement}). 
In addition, the benchmark provides \textbf{a unified, human-aligned agentic physical design workflow framework} that enables closed-loop evaluation of holistic physical design in realistic EDA environments (\textbf{by preserving privacy when integrating commercial EDA tools, PDKs, and frontier LLMs/VLMs}). 
Experiments on 11 state-of-the-art models reveal that while modern LLMs/VLMs perform competitively on conceptual tasks (e.g., GPT-5.5 achieves 73.3\% on root-cause analysis), they remain substantially limited in tool-centric execution and long-horizon, multi-stage reasoning.
Our studies further show that human-skill-enhanced agentic workflows significantly improve end-to-end physical design performance.
Specifically, \textbf{with human instructions, our experiments demonstrate the superior capabilities of VLM in macro placement beyond state-of-the-art analytical and reinforcement learning-based methods.}
\textsc{PDAgent-Bench} establishes a standardized, reproducible, and realistic evaluation framework for advancing LLM/VLM-driven holistic physical design automation{\footnote{All findings and conclusions presented in this work are based solely on the authors’ independent experiments and analyses, and do not reflect or imply any views, opinions, or endorsements from industry.}}.
To ensure full reproducibility, we will release \textsc{PDAgent-Bench} together with its agentic workflow framework, instantiated on open-source PDKs (e.g.,~Nangate45) and open EDA tools, at \url{https://github.com/fenggev56/PDAGENT-BENCH}.


\end{abstract}

\input{Intro}

\input{background}

\input{framework}

\input{experiment}

\input{conclusion}

\bibliography{reference}
\bibliographystyle{abbrvnat}

\input{appendix}



\newpage

\end{document}

%% file: Intro.tex
\section{Introduction}

Very Large-Scale Integrated Circuits (VLSI) are foundational hardware to power various modern technological advances, such as AI, scientific computing, data centers, and more~\cite{vlsi_modern_system, kahng_vlsi}.
The explosive growth of applications in these critical sectors is driving an unprecedented demand for VLSI design with higher productivity, better performance, and greater scale.
Yet, conventional VLSI design workflows are increasingly unable to keep pace  (Figure~\ref{fig:workflow}). 
A fundamental limitation lies in the \textbf{inherently iterative and human-in-the-loop design paradigm}, where engineers must repeatedly orchestrate, analyze, and refine design decisions across the entire stack, from front-end synthesis to back-end physical design~\cite{ml_eda}.
This tightly coupled, multi-stage workflow results in substantial engineering overhead, long feedback cycles, and limited scalability, particularly as design complexity continues to grow~\cite{eda_ic}.
Recent advances in large language models (LLMs) and vision-language models (VLMs) have opened a promising direction toward \textit{agentic design automation}, in which LLM/VLM-based agents can iteratively reason, act, and optimize solutions in closed-loop settings.
Specifically, in front-end design, LLMs have demonstrated strong capabilities in generating and debugging hardware code, significantly reducing manual effort~\cite{verilogcoder, codev, chipnemo}. 
For example, NVIDIA’s ChipNeMo~\cite{chipnemo} shows that high-quality RTL (Register-Transfer Level) code can be generated with just a few prompts. 
Importantly, these advances have not emerged in isolation; rather, they have been enabled and sustained by \textbf{well-defined benchmarks and evaluation frameworks}~\cite{verilogEvalv2, rtllmv1, rtllmv2}, which provide standardized tasks, reproducible protocols, and quantitative metrics for assessing progress.

Yet, \textbf{physical design remains largely underexplored with LLMs/VLMs}, despite its decisive role in determining a chip's final power, performance, and area (PPA). 
The primary barrier to progress is the \textbf{lack of standardized benchmarks} that can capture the full complexity of physical design and enable systematic evaluation of LLM/VLM-based agentic physical design workflows.
It is important to note that physical design presents fundamentally different challenges from front-end tasks: \textbf{physical design involves complex geometric constraints, strict technology design rules, multi-tool orchestration, and long-latency feedback loops}.
Thus, it requires integrating diverse cognitive capabilities: understanding design rules, interpreting tool-generated reports, diagnosing violations, generating tool scripts, and coordinating iterative optimization across stages. 
Existing benchmarks fail to capture this complexity. 
Prior works such as ChiPBench~\citep{chipbench} focus on isolated subproblems (e.g., macro placement) with conventional algorithms, while ChatEDA-Bench~\citep{chateda} and iScript-Bench~\citep{iscript} evaluate narrow capabilities such as script generation. 
As a result, \textbf{current benchmarks are fragmented and disconnected from end-to-end workflows}, leaving them insufficient to assess whether LLM/VLM agents can effectively support realistic physical design.

In this work, we seek to evaluate the capabilities of LLM/VLM-based agents that are engineered to enable \textit{holistic physical design} in realistic EDA environments. 
Formally, this pursuit requires assessing an agent’s ability to (i) perform various types of tasks throughout the design stack, (ii) operate within multi-stage tool-driven workflows, and (iii) optimize design objectives (e.g., PPA) through iterative interaction with EDA tools.
Achieving the goal requires a benchmark that satisfies several key desiderata.
\textbf{(D1) Coverage.} It must span the full spectrum of physical design capabilities, beyond isolated tasks.
\textbf{(D2) Workflow Fidelity.} Evaluation must reflect realistic multi-stage workflows with tool-in-the-loop feedback.
\textbf{(D3) Standardization.} Tasks, inputs, and metrics must be reproducible and comparable across models.
\textbf{(D4) Compositionality.} The benchmark should capture interactions between capabilities (e.g., report understanding → debugging → script generation → tool use).
\textbf{(D5) Diagnostic Ability.} It should enable fine-grained analysis of failure modes and capability gaps.

To this end, we introduce \textbf{\textsc{PDAgent-Bench}}, a first-of-its-kind \textbf{\underline{bench}mark for holistic and reproducible evaluation of LLM-based \underline{agent}ic \underline{p}hysical \underline{d}esign}. 
Our benchmark integrates both task-level evaluation and workflow-level execution.
\textbf{Benchmark Design.}
\textsc{PDAgent-Bench} consists of a \textbf{multi-dimensional benchmark suite} with 353 curated tasks spanning five capability dimensions: (i) foundational knowledge understanding, (ii) report comprehension, (iii) root-cause analysis, (iv) script generation, and (v) full-flow orchestration. 
Each task is designed to reflect realistic scenarios in physical design and is paired with standardized evaluation criteria.
\textbf{Evaluation Framework.}
We further develop a \textbf{unified human-aligned agentic design workflow framework} that enables closed-loop interaction between LLM/VLM agents and various EDA tools. 
This framework allows agents to iteratively \textbf{place macros}, generate actions (e.g., scripts), receive tool feedback (e.g., timing or congestion reports), and refine decisions, thereby enabling evaluation under realistic design loops rather than static prompts.
Using this framework, we conduct a comprehensive evaluation on 11 state-of-the-art LLMs/VLMs.
Our results provide the first systematic characterization of LLM/VLM capabilities in physical design, revealing critical limitations in long-horizon reasoning, tool interaction, and cross-stage coordination.

Our key contributions are summarized as follows:
\textbf{(i)} We propose \textbf{\textsc{PDAgent-Bench}}, the first benchmark for \textbf{end-to-end agentic VLSI physical design workflows}, enabling standardized and reproducible evaluation.
\textbf{(ii)} We construct a \textbf{353-task benchmark suite} covering five key capability dimensions, bridging isolated tasks and holistic workflows.
\textbf{(iii)} We develop a \textbf{unified agent-in-the-loop framework} for closed-loop evaluation with EDA tools by plugging in existing LLMs/VLMs.
\textbf{(iv)} We present a \textbf{systematic empirical study} on 11 LLMs/VLMs, \textbf{showing great potential of VLMs in macro placement and identifying key capability gaps, failure modes, and critical insights for domain-specific model development}.

%% file: background.tex
\section{Background and Related Work}
\label{sec: bg}


\begin{figure*}[!t]
\centering
\includegraphics[width=0.75\linewidth]{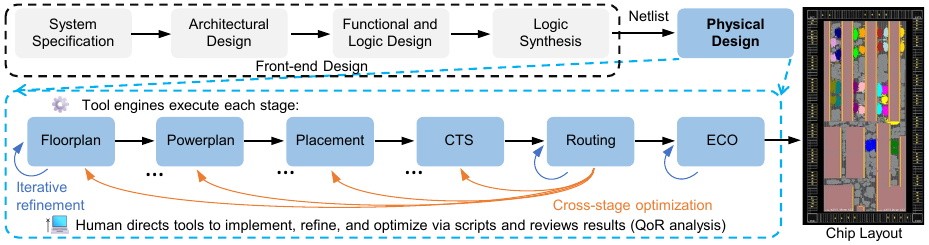}
\caption{Modern VLSI design workflow. It spans from system specification through physical design, with physical design encompassing floorplan through engineering change order (ECO). Each stage involves iterative refinement within the stage (blue) and may trigger cross-stage optimization that loops back to earlier stages (orange). EDA tool engines execute each stage, while human engineers direct the tools via scripts and constraints, and review results with quality-of-results (QoR) analysis.}
\label{fig:workflow}
\end{figure*}

\textbf{VLSI Design Workflow.}
A typical VLSI design workflow used by human designers is shown in Figure~\ref{fig:workflow}, which has two phases: \textit{front-end design} and \textit{physical design}. 
The front-end phase includes architectural design, functional and logic design (hardware coding), and logic synthesis, where high-level functional specifications are translated into a gate-level netlist.
This netlist serves as an input to the physical design phase, which generates a manufacturable layout.
Physical design is a \textbf{multi-stage, tightly coupled optimization process} comprising several key steps~\cite{eda_ic}.
\textbf{Floorplanning} defines the spatial organization of macros and functional blocks, including core area, I/O pin locations, and macro placement, often requiring substantial manual guidance due to limited automation in current tools.
\textbf{Power planning} constructs a hierarchical power delivery network (PDN) (e.g., rings, stripes, meshes, vias) to satisfy IR-drop and electromigration constraints while minimizing routing overhead.
\textbf{Placement} assigns legal locations to standard cells, optimizing objectives such as wirelength, congestion, timing, and density.
\textbf{Clock Tree Synthesis (CTS)} builds a buffered clock network to distribute the clock signal with controlled skew, latency, and transition characteristics.
\textbf{Routing} establishes signal connectivity across metal layers under strict design-rule constraints, balancing wirelength, congestion, and manufacturability.
\textbf{Engineering Change Order (ECO)} performs localized post-route modifications, including cell resizing, buffer insertion, and metal-only fixes, to resolve residual timing violations and design-rule errors without disrupting the global layout.
Although each stage is supported by specialized EDA tools, achieving high-quality designs still relies heavily on \textbf{human experts to orchestrate the tools in the design workflow}. 
Importantly, these stages are \textbf{strongly interdependent}: early decisions (e.g., floorplanning) constrain downstream feasibility; placement impacts congestion and timing closure; and both influence the effectiveness of CTS, routing, and ECO. As a result, physical design requires \textbf{iterative, cross-stage refinement under long feedback loops}, making it a complex, feedback-driven process that is difficult to automate and scale.

\textbf{Agentic Modeling and Benchmarking.}
LLM/VLM-based agents are well-suited for VLSI physical design due to two key properties in the physical design workflow.
\textbf{(i) Script-driven, tool-mediated interaction.} Physical design stages are executed via parameterized tool commands (e.g., Tcl scripts), which encode key decisions such as PDN configuration, placement constraints, and CTS strategies. 
Consequently, the workflow can be abstracted as a sequence of \textit{state-action-feedback} interactions, where an agent observes design states (e.g., reports, constraints), generates executable actions (scripts), and receives feedback from tools. 
This aligns closely with the LLM/VLM agent paradigm of reasoning, tool invocation, and iterative refinement.
\textbf{(ii) Structural regularity and compositional skill reuse.} Physical design exhibits strong regularity across designs within the same circuit class (e.g., CPUs, accelerators). 
Common patterns arise in floorplanning templates, PDN topologies, placement strategies, and timing closure heuristics.
In practice, expert designers leverage this regularity by reusing and adapting prior solutions rather than solving each instance from scratch. 
This suggests that effective automation requires not only reasoning, but also \textbf{retrieval and composition of reusable skills} (e.g., script templates, optimization procedures). 
Such properties make physical design a natural setting for evaluating \textbf{generalization, compositionality, and skill reuse} in LLM agents.
Together, these properties make physical design a natural setting for \textbf{closed-loop, tool-integrated evaluation} of LLM agents, motivating the need for a holistic benchmark for agentic physical design.

\input{tab_comparison_benchmark}

\textbf{Related Work.}
Recently, several benchmarks~\citep{rtllmv1, rtllmv2, verilogEvalv2, circuitnet3, chipbench, chateda, iscript, ispd2019, ml_in_eda} have been proposed to evaluate LLM/VLM-based agents and conventional optimization algorithms for VLSI design, as summarized in Table~\ref{tab:benchmark-comparison}. 
VerilogEval v2~\citep{verilogEvalv2} is restricted to RTL code generation, while CircuitNet 3.0~\citep{circuitnet3} is confined to early-stage timing and power prediction. 
ChipBench~\citep{chipbench} focuses on macro-placement algorithms (analytical methods + reinforcement learning), and ChatEDA-Bench~\citep{chateda} and iScript-Bench~\citep{iscript} primarily evaluate EDA script generation.
Despite these contributions, existing benchmarks share a key limitation: they evaluate \textbf{isolated capabilities} rather than \textbf{end-to-end workflows}.
Beyond benchmarks, recent agent methods~\cite{ORFS-agent, OpenROAD_Agent, schemav2, EDA-schema} such as ORFS-agent~\citep{ORFS-agent} and OpenROAD Agent~\citep{OpenROAD_Agent} demonstrate closed-loop interaction with EDA tools, iteratively tuning flow parameters and generating self-correcting scripts from tool feedback, precisely the capabilities that existing benchmarks fail to capture.
However, none of the current physical design benchmarks capture the closed-loop interaction between agents and EDA tools, nor do they assess critical capabilities such as report comprehension, diagnostic reasoning, and iterative refinement across stages.
As a result, they fail to reflect the \textbf{feedback-driven, multi-stage nature} of real-world physical design, leaving a significant gap in the evaluation of whether LLM/VLM agents can operate effectively in practical settings.
\textsc{PDAgent-Bench} addresses this critical gap by evaluating all these dimensions jointly (Table~\ref{tab:benchmark-comparison}).

%% file: tab_comparison_benchmark.tex
\begin{table*}[t]
\centering
\caption{Comparison of \textsc{PDAgent-Bench} with other circuit benchmarks.}
\label{tab:benchmark-comparison}
\setlength{\tabcolsep}{4pt}
\renewcommand{\arraystretch}{1.15}
\scriptsize
\scalebox{0.96}{
\begin{tabular}{lccccc}
\toprule
\textbf{Benchmark Features} & {VerilogEval V2~\citep{verilogEvalv2}} & {CircuitNet 3.0~\citep{circuitnet3}} & {ChiPBench~\citep{chipbench}} & {ChatEDA-Bench~\citep{chateda}} & \textbf{\textsc{PDAgent-Bench} (Ours)} \\
\midrule
\textbf{Open Source}                              & \checkmark & \checkmark & \checkmark & \checkmark & \checkmark \\
\textbf{RTL-to-GDS Flow}                          & $\times$   & $\times$   & \checkmark & \checkmark & \checkmark \\
\textbf{Industrial Setting}\textsuperscript{\dag} & $\times$   & \checkmark & $\times$   & $\times$   & \checkmark \\
\textbf{Domain Knowledge}                         & $\times$   & $\times$   & $\times$   & $\times$   & \checkmark \\
\textbf{Report Analysis}                          & $\times$   & $\times$   & $\times$   & $\times$   & \checkmark \\
\textbf{Root Cause Analysis}                      & $\times$   & $\times$   & $\times$   & $\times$   & \checkmark \\
\textbf{Static Timing Analysis}                   & $\times$   & $\times$   & $\times$   & $\times$   & \checkmark \\
\textbf{\# of Data}                           & 153        & 8{,}659    & 20         & 50         & 353        \\
\midrule
\textbf{Target Tasks}                             & RTL        & Early-Stage      & Macro        & Script     & \textbf{Comprehensive}\   \\
                                                  & Generation & Timing/Power     & Placement    & Generation & \textbf{Physical Design (including}  \\
                                                  &            & Prediction       & Optimization &            & \textbf{ macro placement)}             \\
\bottomrule
\end{tabular}}
\\[4pt]
{\footnotesize \textsuperscript{\dag}\textbf{Industrial Setting}: uses commercial EDA tools on advanced technology nodes ($\leq$ 28\,nm).}
\end{table*}

%% file: framework.tex
\section{\textsc{PDAgent-Bench}}


\textbf{\textsc{PDAgent-Bench} consists of two complementary components:}
(i) a \textit{multi-dimensional benchmark dataset} that captures the breadth and complexity of physical design tasks, and
(ii) a \textit{tool-integrated, multi-agent design workflow framework} that enables evaluation under realistic EDA interactions.
Together, these components support \textbf{holistic, standardized, and reproducible evaluation} of LLM/VLM agents in holistic physical design.

\subsection{Benchmark Development}

\textbf{Overview.} To systematically evaluate LLM/VLM capabilities in physical design, we construct a benchmark of 353 curated problems (Table~\ref{tab:bench-overview}).
The dataset spans a spectrum from foundational knowledge questions to production-oriented tasks, including report comprehension, root-cause analysis, optimization reasoning, and script generation. 
This design aims to capture the core competencies required of a physical-design engineer while enabling fine-grained analysis of model performance.

{\textbf{Data Selection Criteria}.} Each problem in our benchmark dataset is selected to satisfy the following principles.
\textbf{(C1) Coverage and diversity.} Tasks span multiple stages of the physical design flow and require both domain knowledge and structured reasoning.
\textbf{(C2) Ground-truth fidelity.} Each instance is paired with a detailed reference solution to enable precise evaluation and error analysis.
\textbf{(C3) Multi-modality.} Problems incorporate text, scripts, and visual elements (e.g., floorplans, design rule check (DRC) maps), reflecting real-world inputs.
\textbf{(C4) Leakage resistance.} Instances are curated to minimize overlap with publicly available data, reducing the risk of memorization-based performance.
\textbf{(C5) Realistic difficulty.} Tasks extend beyond simple question-answering (QA) to include complex operations such as timing report interpretation and full-flow reasoning.

\input{dataset_overview}

{\textbf{Data Representation and Preprocessing}.}
The raw dataset comprises 3 modalities, including scripts, free-form text, and images.
To enable consistent processing and evaluation, we normalize all entries into a \textbf{unified JSON schema}, ensuring structured representation and efficient parsing.
For \textbf{script generation tasks}, we provide a comprehensive set of acceptable solutions, accounting for multiple valid command variants due to tool-version differences, alternative syntactic forms, and synonymous expressions. 
For \textbf{question-answering tasks}, we design detailed evaluation rubrics with clearly defined scoring criteria and key points, enabling reliable and consistent \textbf{LLM-as-judge} assessment.

\input{dataset_circuit_overview}

{\textbf{Benchmark}.}
Following the criteria and preprocessing strategies, we construct \textsc{PDAgent-Bench} benchmark dataset, summarized in Table~\ref{tab:bench-overview}. 
The benchmark comprises two complementary components. 
\textbf{First, foundational suite (90 instances)}. We curate 90 foundational questions that cover the full physical design flow, from initialization to tape-out, ensuring broad coverage of core concepts and design stages. 
\textbf{Second, production-oriented suite (263 instances)}. 
To evaluate performance on real-world, production-oriented tasks, we include 263 additional instances derived from commonly used implementation scripts, timing reports, and error/violation reports collected from our projects, whose results have been published at top-tier solid-state circuit design conferences including ISSCC~\cite{isscc_name}, VLSI Symposium~\cite{vlsi_name}, A-SSCC~\cite{asscc_name}, and CICC~\cite{cicc_name}. 
The script generation spans the major commercial EDA tools used across the modern physical-design flow.
To ensure quality and reliability, all questions and reference answers are independently reviewed by three domain experts.
In addition, all script-based solutions are validated through execution, confirming functional correctness and practical usability within a simulation environment.
We organize the benchmark into three evaluation suites.
The \textbf{Physical Design Understanding suite (133 instances)} evaluates foundational concepts, EDA report comprehension, root-cause analysis, and static timing analysis. 
The \textbf{Script Generation suite (210 instances)} tests executable Tcl generation across the major commercial EDA tools. 
The \textbf{Full-Flow Implementation suite} (Table~\ref{tab:bench-designs}) drives a complete RTL-to-GDSII (physical layout) flow on 10 designs, ranging from a 176-cell FSM (finite state machine) to a 1.5M-cell NVDLA accelerator, graded on placement quality, timing closure, DRC (design rule check)/LVS (layout vs. schematic) cleanliness, and PPA. 
As summarized in Table~\ref{tab:bench-overview}, \textbf{our benchmark is designed to probe the key capabilities required of agents, spanning four core aspects: planning, reasoning, tool use, and perception}~\cite{agent_capability}.

\subsection{Agentic Design Workflow Framework}
\label{sec:framework}

\begin{figure*}[!t]
\centering
\includegraphics[width=1.0\linewidth]{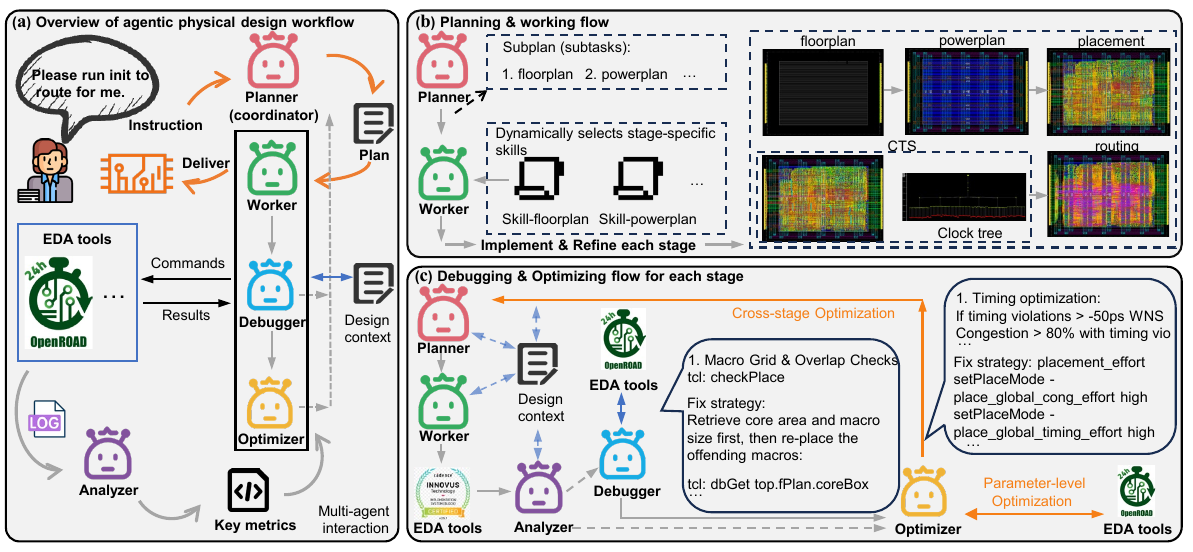}
\caption{(a) {Overview of the PDAgent framework.} Five specialized agents (Planner, Worker, Debugger, Optimizer, and Analyzer) coordinate to translate user instructions into a complete physical-design flow. (b) The Planner decomposes the task into stage-wise subplans, and the Worker invokes stage-specific skills for each stage, advancing only after debugging and optimization are complete. (c) The Optimizer and Debugger refine the design through targeted fix strategies until convergence.}
\label{fig:openlayout_agent}
\end{figure*}

\textbf{Overview.}
To enable systematic benchmarking of human-aligned agentic workflows for VLSI physical design, we develop \textbf{\textsc{PDAgent}}, a unified framework for constructing and evaluating LLM/VLM-based agents in closed-loop, tool-integrated environments (Figure~\ref{fig:openlayout_agent}(a)). 
The framework defines standardized \textbf{agent roles, interfaces, and coordination mechanisms}, allowing researchers to instantiate and compare diverse agentic workflows under consistent conditions. 
It is designed to be \textbf{modular and extensible}, supporting plug-and-play integration of different models (e.g., open-source or proprietary LLMs/VLMs) and customization for varying design objectives.
In particular, our framework adopts a privacy-preserving approach to integrate commercial EDA tools, PDKs, and LLMs/VLMs through a local environment (Section~\ref{sec: privacy-preserving}). This corresponds to partially addressing interoperability and deployment issues due to the multi-party involvement, an important domain-specific gap raised in the recent Cognitive Silicon Design Workshop (CSDW)~\cite{CSDW}. 

Given a gate-level netlist, technology constraints, and design targets, \textsc{PDAgent} orchestrates agents to iteratively produce a DRC-clean, timing-closed layout. 
The workflow follows the natural structure of physical design, comprising four phases, \textit{planning, implementation, debugging, and optimization}, which form a closed loop for progressive refinement.
Specifically,  \textsc{PDAgent} comprises a Planner Agent, Worker Agent, Debugger Agent, and Optimizer Agent, augmented by an Analyzer Agent that extracts key metrics from verbose EDA tool logs. 
We briefly introduce the key design knobs of \textsc{PDAgent} below.
Detailed descriptions of each agent's role are provided in Appendix~\ref{appendix:agent_details}.

\textbf{Agent Roles.}
The \textsc{PDAgent} comprises five agents, each encapsulating a distinct concern:
\textbf{(i)}~\emph{Planner} serves as the central hub, decomposing user intent into a directed acyclic graph (DAG) of subtasks, routing all inter-agent messages, and maintaining global execution;
\textbf{(ii)}~\emph{Worker} interfaces with the EDA tool (e.g., Innovus) via a persistent interactive shell, executing TCL commands and returning structured results, illustrated in Figure~\ref{fig:openlayout_agent}(b). 
Notably, \textbf{macro placement during the floorplan stage is performed directly by the NLMs agents}, which takes the layout image as visual input to guide placement decisions (see Appendix~\ref{sec:agent_design} for details); 
\textbf{(iii)}~\emph{Analyzer} extracts quantitative metrics, timing, for example, worst negative slack (WNS), physical (DRC, utilization), and power, from EDA logs;
\textbf{(iv)}~\emph{Debugger} diagnoses failures through LLM/VLM-powered root-cause analysis;
\textbf{(v)}~\emph{Optimizer} performs parameter tuning guided by metric trajectories and domain knowledge, as shown in Figure~\ref{fig:openlayout_agent}(c).

\textbf{Hub-and-Spoke Communication.}
All inter-agent messages are routed through the Planning Agent, which 
maintains full observability of the system state and enforces execution 
ordering. The sole exception is \emph{tool interaction}: any agent may 
directly invoke EDA tool commands via a synchronous 
\texttt{execute\_tool} interface, bypassing hub routing to avoid 
latency overhead for frequent tool queries. This hybrid topology 
balances centralized coordination with efficient tool interaction.

\textbf{Skill Packs.}
Building on the inherent skill-driven physical design workflow introduced in Section~\ref{sec: bg}, the agentic framework is \emph{grounded} in a curated skill library that encodes the procedural knowledge of senior engineers, yielding a skill-enhanced LLM/VLM-based agentic framework. 
Skills encode procedural knowledge, executable scripts together with the rationale behind parameter choices and PPA optimization strategies, distilled from senior engineers. 
This kind of tacit, experience-driven expertise is what traditional algorithmic approaches cannot capture, and what motivates an LLM/VLM-based agentic framework.
Domain knowledge is organized into \emph{skill packs}, self-contained directories structured as \texttt{skills\_\{tech\}\_\{tool\}/} (e.g., \texttt{skills\_tsmc28nm\_innovus/}).
Each skill pack contains per-stage skill files that encode expert knowledge: reference TCL scripts, parameter-tuning strategies, debug decision matrices, and cross-stage impact analyses.
Skills are auto-discovered at initialization and resolved dynamically via fuzzy matching, enabling seamless extension to new technology nodes and EDA tools without modifying agent code.

\textbf{Closed-Loop PPA Optimization.}
Aligning with the practical VLSI design, the optimization mechanism operates at two levels: a \textbf{stage-level} loop that progresses through the physical design flow step by step, and a \textbf{parameter-level} loop that explores the design space within each stage. 
When design targets are unmet, the agent employs a \textbf{tree-of-thought exploration strategy}, generating and evaluating multiple candidate actions before selecting the most promising trajectory. Iteration continues until convergence or a predefined budget is reached.

%% file: dataset_overview.tex
\begin{table*}[!t]
  \centering
  \caption{Overview of benchmark categories used to evaluate agents on physical design tasks. Report
Comprehension, Root Cause Analysis, and Static Timing Analysis are multi-modal tasks that pair text with EDA reports, layouts,
or timing diagrams. The targeted agent capabilities are denoted by: \ding{172} Reasoning, \ding{173} Planning, \ding{174} Perception, \ding{175} Tool Use.}
  \label{tab:bench-overview}
  \scriptsize
  \begin{tabular}{l l c p{5.5cm} c}
    \toprule
    \textbf{Task} & \textbf{Category} & \textbf{\#} & \textbf{Description} & \textbf{Agent Capability} \\
    \midrule
    \multirow{4}{*}[-2.0\baselineskip]{\begin{tabular}[c]{@{}l@{}}Physical Design\\Understanding\end{tabular}}
      & Foundational Knowledge   & 90 & Conceptual questions on physical-design and EDA fundamentals (timing, power, placement, routing, DRC/LVS). & \ding{172}, \ding{173} \\
      & Report Comprehension     & 11 & Interpreting EDA tool reports (timing, congestion, power, utilization) to extract metrics and identify issues. & \ding{172}, \ding{174} \\
      & Root Cause Analysis      & 21 & Diagnosing the underlying cause of PPA or convergence problems and proposing optimization directions. & \ding{172}, \ding{174} \\
      & Static Timing Analysis   & 11 & Path-level reasoning over setup/hold violations, slack, clock skew, and timing exceptions. & \ding{172}, \ding{174} \\
    \midrule
    \multirow{4}{*}[-1.0\baselineskip]{\begin{tabular}[c]{@{}l@{}}Script\\Generation\end{tabular}}
      & P\&R Vendor 2                  & 90 & Generating Cadence Innovus TCL scripts for place-and-route flow steps. & \ding{175} \\
      & P\&R Vendor 1                     & 88 & Generating Synopsys IC Compiler II TCL scripts for backend implementation. & \ding{175} \\
      & ECO                      & 10 & Generating engineering-change-order scripts for post-route fixes (timing, DRC, functional). & \ding{175} \\
      & FM (Formal Verification) & 22 & Generating Synopsys Formality scripts for RTL-vs-netlist equivalence checking. & \ding{175} \\
    \midrule
    \begin{tabular}[c]{@{}l@{}}Full Flow\\Implementation\end{tabular}
      & \begin{tabular}[c]{@{}l@{}}End-to-End \textbf{placement}\\and \textbf{route (PnR)}\end{tabular}
                                  & 10 & Driving a complete RTL-to-GDSII flow on the design suite below, evaluated on PPA, timing closure, and DRC/LVS cleanliness. & \begin{tabular}[c]{@{}c@{}}\ding{172}, \ding{173},\\ \ding{174}, \ding{175}\end{tabular} \\
    \bottomrule
  \end{tabular}
\end{table*}

%% file: dataset_circuit_overview.tex
\begin{table*}[!t]
  \centering
  \caption{Benchmark designs used in the full-flow implementation. We report the cell count after synthesis, the target clock period, and the utilization target.}
  \label{tab:bench-designs}
  \small
  \setlength{\tabcolsep}{4pt}
  \scalebox{0.92}{
  \begin{tabular}{l l r c r}
    \toprule
    \textbf{Design} & \textbf{Description} & \textbf{\#Cells} & \textbf{Target Clk (ns)} & \textbf{Util.} \\
    \midrule
    TinyRISCV~\cite{tinyriscv}            & Compact RISC-V core (small control-logic design).           & 15{,}012    & 5.0 & 0.65 \\
    Open910~\cite{910}              & Open-source application-class processor.                    & 754{,}981   & 1.5 & 0.60 \\
    Systolic array~\cite{systolic_array_rtl}       & Matrix-multiply accelerator (datapath-heavy).               & 41{,}198    & 3.4 & 0.50 \\
    16b pipelined mult.~\cite{ethmac_opencores}  \ & 16-bit pipelined multiplier (arithmetic datapath).          & 342         & 2.6 & 0.55 \\
    AES-256~\cite{aes}              & AES-256 cryptographic core.                                 & 15{,}516    & 3.5 & 0.60 \\
    NVDLA-small~\cite{nvdla}          & NVIDIA Deep Learning Accelerator, small config.             & 270{,}072   & 2.0 & 0.60 \\
    NVDLA-large~\cite{nvdla}          & NVIDIA Deep Learning Accelerator, large config.             & 1{,}478{,}865 & 2.5 & 0.60 \\
    UART controller~\cite{uart}      & Serial communication peripheral.                            & 368         & 3.4 & 0.65 \\
    Elevator FSM~\cite{ethmac_opencores}         & Finite-state-machine control design.                        & 176         & 2.5 & 0.70 \\
    Ethernet MAC~\cite{ethmac_opencores}         & Ethernet media access controller.                           & 35{,}172    & 10.0 & 0.60 \\
    \bottomrule
  \end{tabular} }
\end{table*}

%% file: experiment.tex
\section{Experiments and Results}

This section presents comprehensive evaluations of various LLM/VLM-based agents on VLSI physical design tasks by using \textsc{PDAgent-Bench}.
We first describe the experimental setup, then report the main results, and finally summarize key findings and insights.

\subsection{Experimental Setup}

\textbf{Models and Computing Platforms.} We evaluate 11 LLMs/VLMs with \textsc{PDAgent-Bench}, including 6 proprietary and 5 open-source models. 
The proprietary models are GPT-5-mini, GPT-5.1, GPT-5.4-mini, GPT-5.5~\cite{gpt5}, Claude Sonnet 4.6~\cite{claude-sonnet-4.6}, and Claude Opus 4.7~\cite{claude-opus4.7}. 
The open-source models are Qwen3-VL-8B, Qwen3-VL-30B~\cite{qwen3-vl}, Qwen3-8B~\cite{qwen3}, DeepSeek-Chat, and DeepSeek-Reasoner~\cite{deepseek}.
All experiments are done with 12 NVIDIA A6000 Ada GPUs and an AMD EPYC 7313 CPU.

\textbf{EDA Tools and Semiconductor Nodes}. \textbf{The primary goal of our work is to benchmark the capabilities of LLMs/VLMs for VLSI physical design, not to compare or rank different EDA tools}. 
Commercial EDA tools are used only as part of the evaluation infrastructure to assess the designs guided by LLMs/VLMs.
\textbf{We use a privacy-preserving framework to integrate commercial EDA tools and LLMs/VLMs without exposing the initial scripts (Section~\ref{sec: privacy-preserving})}.
We use Cadence Innovus 22~\cite{innovus}, Synopsys IC Compiler II
(ICC2) 2022~\cite{icc2}, Synopsys PrimeTime (PT) 2022~\cite{pt}, and Synopsys Formal (FM) 2022~\cite{fm} as our commercial EDA tools for implementation and signoff. 
Command-level differences between tool versions are minimal; we have validated that our generated scripts are compatible with versions 2018 and 2020.
For experiments using commercial EDA tools, we target the TSMC 28-nm process; for the open-source flow, we adopt OpenROAD with the Nangate 45-nm node.

\textbf{Evaluation Metrics and Sampling Configuration.} 
We report \textbf{pass@1} and \textbf{pass@5} for \textit{Physical Design Understanding} and \textit{Script Generation} tasks.
For free-response foundational knowledge, root-cause analysis, and report comprehension tasks, we generate five responses per query and score each against the reference answer using predefined rubrics. 
Pass@1 is obtained from a single response generated with temperature 0 (deterministic decoding).
Pass@5 is computed by generating five responses with temperature 0.8 (top-p = 0.95) and reporting the best rubric score among them.
For full-flow implementation tasks, we report \textbf{stage-specific design metrics}, including WNS, placement density, and the number of DRC violations, which directly reflect design quality and PPA outcomes.
For open-source models, we use deterministic decoding (temperature = 0) for pass@1, and stochastic sampling (temperature = 0.8, top-p = 0.95) for pass@5. 
For proprietary APIs that do not expose a temperature parameter, we sample five responses per query using the default configuration.

\textbf{LLM-as-Judge and Human Evaluation.} To enable affordable, large-scale qualitative analysis of model outputs, we adopt the widely used LLM-as-Judge protocol~\cite{llm_as_judge} with Claude Sonnet~4.6 as the judge. 
Full prompts are provided in Appendix~\ref{system_prompt}.
To validate our LLM-Judge protocol, we recruit three senior physical-design engineers to independently score the model outputs on Basic, RootOpt, and Report using the same rubric provided to the LLM judge. 
We report the mean score across the three engineers as the human-evaluation result in Table~\ref{tab:bench-models}.


\subsection{Experimental Results}

\begin{figure*}[!t]
\centering
\includegraphics[width=1.0\linewidth]{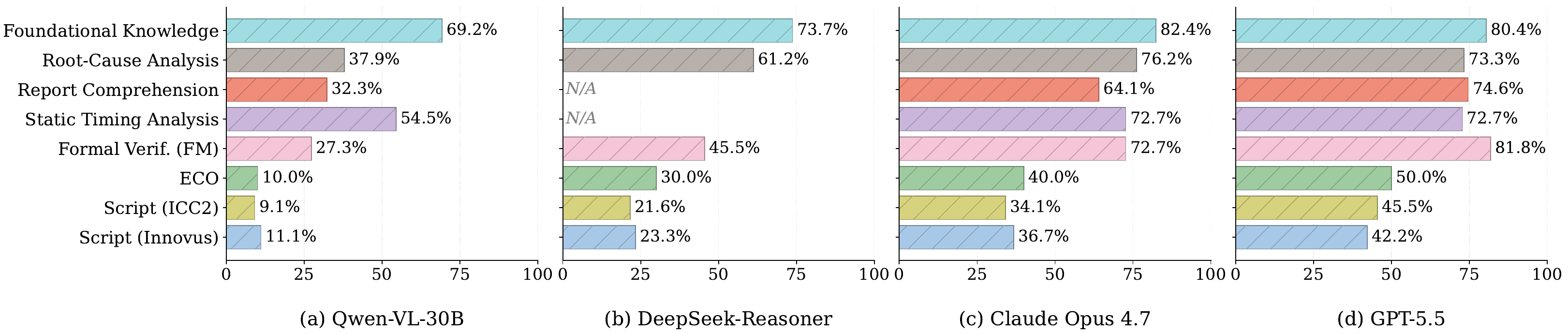}
\caption{Per-task accuracy (\%) of four LLMs across eight evaluation categories. Frontier models (Claude Opus 4.7, GPT-5.5) lead across nearly all tasks, while all models struggle on script generation and ECO. \textit{N/A} denotes tasks requiring multi-modal input, which the model (LLM) does not support.}
\label{fig:pass1_bar}
\end{figure*}

\textbf{Physical Design Understanding and Script Generation.}
Table~\ref{tab:bench-models} summarizes the key results, with per-task pass@1 of the top four models illustrated in Figure~\ref{fig:pass1_bar}. 
GPT-5.5 achieves state-of-the-art performance on five of eight tasks, attaining 45.5\% and 42.2\% pass@1 on ICC2 and Innovus script generation, respectively, representing up to a 4$\times$ improvement over the strongest open-source model (Qwen3-VL-30B, at 9.1\% and 11.1\%).
Despite this progress, all models remain below 50\% pass@1 on script generation and 60\% on ECO tasks, highlighting the difficulty of producing executable, tool-specific Tcl code and the limited coverage of EDA-specific data in public training corpora.
In contrast, foundational knowledge emerges as the most tractable category: Claude Opus 4.7 achieves the highest performance at 82.4\%, while even open-source models reach competitive levels (e.g., Qwen3-VL-30B at 69.2\%), suggesting that textbook-level physical design knowledge is well represented in pretraining data.
For reasoning-intensive tasks, including root-cause analysis and report comprehension, GPT-5.5 exceeds 70\% pass@1, whereas open-source models lag by 20$\sim$40 percentage points. 
This substantial gap indicates that complex reasoning and multi-step diagnosis are key differentiators between proprietary and open models in physical design settings.
Human scores (parentheses in Table~\ref{tab:bench-models}) track LLM-Judge scores within 2$\sim$6 points, demonstrating that LLM-as-Judge scoring is a reliable proxy for human evaluation on these tasks.

\input{Table_main_results}


\textbf{Full-Flow Implementation.}
Table~\ref{tab:partial_ff_results} reports partial full-flow implementation results of \textsc{PDAgent} on three representative designs from the PDAgent-Bench full-flow suite (i.e., TinyRISCV, AES-256, and Ethernet MAC); complete results and an ablation against an LLM/VLM agent without our skill library are provided in the Appendix~\ref{app:whole_result_fullflow}. 
The skill-free baseline fails at early stages such as initialization and floorplanning, consistent with the weak script-generation performance reported in Table~\ref{tab:bench-models}. 
In contrast, \textsc{PDAgent} closes timing (positive WNS) and resolves all DRC violations after RouteOpt across the three designs, demonstrating that our agent can drive a complete RTL-to-GDSII flow on real industrial designs in TSMC 28 nm.

\textbf{Macro Placement.} The macro placement is the grand challenge in the VLSI physical design workflow, which still lacks effective algorithms. 
We thus evaluate VLM agents enhanced by Skills in our framework on this specific design stage.
We evaluate them on five ChiPBench designs against seven learning-based and analytical placers: WireMask-EA, simulated annealing (SA), MaskPlace, ChiPFormer, DREAMPlace, AutoDMP, and OpenROAD.
The detailed results are reported in Table~\ref{tab:macro_all} (Appendix~\ref{sec: macroplacement}), demonstrating that our flow, driven by an 8B VLM backbone, can effectively handle complex macro placement and match or outperform specialized macro placers in timing closure.

\input{partial_full_flow}

\subsection{Key Findings and Insights}
\label{sec:findings}

\textbf{Key Findings.} We analyze the quantitative results alongside error cases (Appendix~\ref{app:error_case_analysis}) and summarize the findings, revealing the main challenges and directions for improving LLM-based agentic workflows in physical design.
\textbf{Finding 1: Failures are predominantly syntactic rather than conceptual.} Approximately $92\%$ of script-generation failures are near-misses, where models select the correct command (e.g., in Cadence Innovus) but produce incorrect, missing, or inconsistent arguments.
This suggests that LLMs largely capture the \textit{procedural structure} of physical design workflows, but lack fidelity in tool-specific syntax. 
Bridging this gap requires explicit \textbf{syntax grounding} or retrieval over tool documentation.
\textbf{Finding 2: Strong performance on foundational knowledge.} On the foundational benchmark, GPT-5.5 achieves $80.4\%$ overall across 90 questions, including $83.0\%$ on easy and $81.7\%$ on medium difficulty. 
This indicates that frontier models can reliably answer conceptual and entry-level physical design questions, serving as a strong baseline for knowledge understanding.
\textbf{Finding 3: Perception outpaces decision-making.} While models achieve near-perfect performance on report comprehension ($\sim$9/10), performance drops significantly ($\sim$5/10) when tasks require root-cause diagnosis or optimization actions. 
This reveals a gap between \textbf{interpreting tool outputs} and \textbf{acting on them}, highlighting the need for mechanisms that translate observations into executable design decisions.
\textbf{Finding 4: Skill augmentation enables end-to-end execution.} Integrating curated skill libraries substantially improves performance on full-flow tasks (including the challenging macro placement task) by reducing reliance on fragile script generation. 
With access to reusable procedures, agents focus on higher-level reasoning, e.g., navigating PPA trade-offs and iteratively refining designs, enabling more stable closed-loop execution.

\textbf{Insights.} These findings suggest two complementary directions for advancing LLM/VLM-based physical design. 
\emph{(i) Data and training.} Expanding datasets that align reports with root-cause analyses, optimization rationales, and executable scripts can directly address gaps in syntax fidelity and decision-making (\textbf{Findings 1 and 3}).
\emph{(ii) Skill-centric architectures.} Developing richer and more scalable skill libraries, along with efficient retrieval and composition mechanisms, can improve generalization across designs, tools, and technology nodes (\textbf{Finding 4}).
Finally, training domain-adapted models on such curated data would enable \textbf{on-premises deployment}, which is essential for industrial physical design workflows involving proprietary designs and process design kits (PDKs).

\subsection{Limitations and Future Work}

\noindent\textbf{Limitations.} While \textsc{PDAgent-Bench} provides a comprehensive framework for evaluating LLM/VLMs agents in physical design, several limitations remain.
First, \textbf{signoff completeness} is not fully covered. Layout-versus-Schematic (LVS) verification is currently excluded, leaving a critical stage of the design flow unmodeled.
Second, \textbf{automation coverage} is incomplete for certain tasks, particularly DRC fixing that relies on GUI-based interactions or manual inspection, which are difficult to capture within script-driven workflows.
Third, \textbf{dataset scale and diversity} are limited: although the benchmark includes both foundational and production-oriented tasks, it does not yet fully reflect the scale, heterogeneity, and corner cases of industrial-grade designs.

\noindent\textbf{Future Work.} We outline several directions to extend both the benchmark and the agentic design framework. 
\textbf{First}, we plan to expand \textbf{technology coverage} beyond 28-nm to include both advanced nodes (e.g., 16-nm, 7-nm) and mature nodes (e.g., 65-nm, 130-nm), enabling broader evaluation across design regimes.
\textbf{Second}, we will increase \textbf{dataset scale and complexity} by incorporating larger and more diverse designs, improving the benchmark’s ability to stress-test long-horizon reasoning and workflow robustness.
\textbf{Third}, we aim to develop more \textbf{fine-grained, stage-specific capabilities}, including automated macro placement, congestion-aware floorplanning, and enhanced DRC/LVS handling, to better approximate real-world design workflows.
\textbf{Fourth}, we will \textbf{enhance the general capabilities} of our agentic platform by incorporating modules currently absent from the framework, such as long-term memory and self-improvement mechanisms that enable the agent to learn from past successes and failures.
\textbf{Finally}, we plan to improve \textbf{evaluation fidelity} by incorporating additional signoff metrics and richer feedback signals, further bridging the gap between benchmark settings and production environments.


%% file: Table_main_results.tex
\begin{table*}[!t]
\centering
\caption{Benchmark results across  Physical Design Understanding and Script Generation. Each cell reports LLM-Judge pass@1 / pass@5; values in parentheses are the corresponding \textit{human-evaluation} scores, available for Foundational knowledge (Foundational), RootOpt, and Report. \textbf{Bold} indicates the best LLM-Judge result per column; \underline{underline} indicates the second best. N/A: model does not support multi-modal input.}
\label{tab:bench-models}
\setlength{\tabcolsep}{3pt}
\renewcommand{\arraystretch}{1.1}
\resizebox{\textwidth}{!}{%
\begin{tabular}{lcccccccccccccccc}
\toprule
\textbf{Model} & \multicolumn{2}{c}{\textbf{Foundational}} & \multicolumn{2}{c}{\textbf{RootOpt}} & \multicolumn{2}{c}{\textbf{Report}} & \multicolumn{2}{c}{\textbf{STA}} & \multicolumn{2}{c}{\textbf{FM}} & \multicolumn{2}{c}{\textbf{ECO}} & \multicolumn{2}{c}{\textbf{P\&R Vendor 1}} & \multicolumn{2}{c}{\textbf{P\&R Vendor 2}} \\
\cmidrule(lr){2-3}\cmidrule(lr){4-5}\cmidrule(lr){6-7}\cmidrule(lr){8-9}\cmidrule(lr){10-11}\cmidrule(lr){12-13}\cmidrule(lr){14-15}\cmidrule(lr){16-17}
 & p@1 & p@5 & p@1 & p@5 & p@1 & p@5 & p@1 & p@5 & p@1 & p@5 & p@1 & p@5 & p@1 & p@5 & p@1 & p@5 \\
\midrule
Qwen-VL-8B & .533\,{\scriptsize(.520)} & .573\,{\scriptsize(.558)} & .245\,{\scriptsize(.228)} & .319\,{\scriptsize(.269)} & .327\,{\scriptsize(.293)} & .464\,{\scriptsize(.421)} & .545 & \underline{.636} & .273 & .455 & .100 & .100 & .068 & .102 & .122 & .133 \\
Qwen-VL-30B & .692\,{\scriptsize(.661)} & .695\,{\scriptsize(.679)} & .379\,{\scriptsize(.368)} & .460\,{\scriptsize(.400)} & .323\,{\scriptsize(.294)} & .427\,{\scriptsize(.415)} & .545 & \underline{.636} & .273 & .545 & .100 & .100 & .091 & .125 & .111 & .178 \\
Qwen3-8B & .543\,{\scriptsize(.488)} & .585\,{\scriptsize(.573)} & .255\,{\scriptsize(.216)} & .286\,{\scriptsize(.263)} & \multicolumn{2}{c}{N/A} & \multicolumn{2}{c}{N/A} & .273 & .500 & .000 & .000 & .080 & .102 & .122 & .144 \\
DeepSeek-Chat & .726\,{\scriptsize(.689)} & .799\,{\scriptsize(.773)} & .486\,{\scriptsize(.434)} & .521\,{\scriptsize(.473)} & \multicolumn{2}{c}{N/A} & \multicolumn{2}{c}{N/A} & .409 & .591 & .300 & \underline{.500} & .340 & .364 & .256 & .356 \\
DeepSeek-Reasoner & .737\,{\scriptsize(.684)} & .806\,{\scriptsize(.788)} & .612\,{\scriptsize(.585)} & .680\,{\scriptsize(.630)} & \multicolumn{2}{c}{N/A} & \multicolumn{2}{c}{N/A} & .455 & .591 & .300 & \underline{.500} & .216 & .352 & .233 & .356 \\
\midrule
GPT-5-mini & .800\,{\scriptsize(.701)} & .820\,{\scriptsize(.790)} & .560\,{\scriptsize(.540)} & .655\,{\scriptsize(.615)} & .655\,{\scriptsize(.607)} & .685\,{\scriptsize(.666)} & .545 & \underline{.636} & .227 & .364 & .300 & .300 & .057 & .114 & .133 & .189 \\
GPT-5.1 & .820\,{\scriptsize(.802)} & \underline{.840}\,{\scriptsize(.830)} & .655\,{\scriptsize(.619)} & .674\,{\scriptsize(.624)} & .655\,{\scriptsize(.604)} & .705\,{\scriptsize(.649)} & .545 & .545 & .545 & .636 & .200 & .200 & .318 & \underline{.432} & .278 & .311 \\
Claude Sonnet 4.6 & \textbf{.826}\,{\scriptsize(.794)} & .835\,{\scriptsize(.821)} & .688\,{\scriptsize(.660)} & .700\,{\scriptsize(.664)} & \underline{.668}\,{\scriptsize(.625)} & \underline{.736}\,{\scriptsize(.717)} & \underline{.636} & \underline{.636} & .636 & .682 & \textbf{.500} & \underline{.500} & \underline{.341} & .352 & .289 & .311 \\
Claude Opus 4.7 & \underline{.824}\,{\scriptsize(.784)} & \textbf{.842}\,{\scriptsize(.829)} & \textbf{.762}\,{\scriptsize(.714)} & \underline{.767}\,{\scriptsize(.731)} & .641\,{\scriptsize(.614)} & .659\,{\scriptsize(.629)} & \textbf{.727} & \textbf{.727} & \underline{.727} & \underline{.727} & \underline{.400} & \underline{.500} & \underline{.341} & .375 & \underline{.367} & \underline{.411} \\
GPT-5.5 & .804\,{\scriptsize(.778)} & .830\,{\scriptsize(.807)} & \underline{.733}\,{\scriptsize(.689)} & \textbf{.774}\,{\scriptsize(.732)} & \textbf{.746}\,{\scriptsize(.720)} & \textbf{.750}\,{\scriptsize(.732)} & \textbf{.727} & \textbf{.727} & \textbf{.818} & \textbf{.864} & \textbf{.500} & \textbf{.600} & \textbf{.455} & \textbf{.580} & \textbf{.422} & \textbf{.511} \\
GPT-5.4-mini & .723\,{\scriptsize(.668)} & .741\,{\scriptsize(.726)} & .545\,{\scriptsize(.502)} & .581\,{\scriptsize(.526)} & .461\,{\scriptsize(.422)} & .514\,{\scriptsize(.487)} & \underline{.636} & \underline{.636} & .591 & \underline{.727} & \textbf{.500} & \underline{.500} & .295 & .398 & .267 & .356 \\
\bottomrule
\end{tabular}%
}
\end{table*}

%% file: partial_full_flow.tex
\begin{table*}[!t]
\centering
\caption{Partial results of \textbf{PDAgent} on the 
PDAgent-Bench Full flow suite (TSMC 28nm); the complete results are reported in Appendix~\ref{app:whole_result_fullflow}. WNS and DRC violations are reported 
at each stage; positive WNS indicates timing closure. Utilization (Util.) 
is reported after floorplanning.}
\label{tab:partial_ff_results}
\scalebox{0.59}{
\begin{tabular}{lllllllll}
\hline
Design                               & Init                                                                                              & Floorplan                                                                        & Powerplan                   & Place                                                                                                  & CTS                                                                                                    & CTSopt                                                                                                & Route                                                                                                   & Routeopt                                                                                               \\ \hline
\multirow{3}{*}{TinyRISCV}           & \multirow{3}{*}{\begin{tabular}[c]{@{}l@{}}5ns, WNS: 1.362ns\\ Setup successfully\end{tabular}}   & \multirow{3}{*}{\begin{tabular}[c]{@{}l@{}}Util: 0.65\\ DRC Vio: 0\end{tabular}} & \multirow{3}{*}{DRC Vio: 0} & \multirow{3}{*}{\begin{tabular}[c]{@{}l@{}}WNS : -0.014ns\\ DRC Vio: 0\\ Density: 0.715\end{tabular}}  & \multirow{3}{*}{\begin{tabular}[c]{@{}l@{}}WNS : -0.127ns\\ DRC Vio: 0\\ Density: 0.7165\end{tabular}} & \multirow{3}{*}{\begin{tabular}[c]{@{}l@{}}WNS : 0.001ns\\ DRC Vio: 0\\ Density: 0.713\end{tabular}}  & \multirow{3}{*}{\begin{tabular}[c]{@{}l@{}}WNS : 0.021ns\\ DRC Vio: 1\\ Density: 0.722\end{tabular}}    & \multirow{3}{*}{\begin{tabular}[c]{@{}l@{}}WNS : 0.021ns\\ DRC Vio: 0\\ Density: 0.722\end{tabular}}   \\
                                     &                                                                                                   &                                                                                  &                             &                                                                                                        &                                                                                                        &                                                                                                       &                                                                                                         &                                                                                                        \\
                                     &                                                                                                   &                                                                                  &                             &                                                                                                        &                                                                                                        &                                                                                                       &                                                                                                         &                                                                                                        \\
\multirow{3}{*}{AES-256}             & \multirow{3}{*}{\begin{tabular}[c]{@{}l@{}}3.5ns, WNS: 1.003ns\\ Setup successfully\end{tabular}} & \multirow{3}{*}{\begin{tabular}[c]{@{}l@{}}Util: 0.60\\ DRC Vio: 0\end{tabular}} & \multirow{3}{*}{DRC Vio: 0} & \multirow{3}{*}{\begin{tabular}[c]{@{}l@{}}WNS : 0.003ns\\ DRC Vio: 0\\ Density: 0.613\end{tabular}}   & \multirow{3}{*}{\begin{tabular}[c]{@{}l@{}}WNS : 0.006ns\\ DRC Vio: 0\\ Density: 0.615\end{tabular}}   & \multirow{3}{*}{\begin{tabular}[c]{@{}l@{}}WNS : 0.006ns\\ DRC Vio: 0\\ Density: 0.615\end{tabular}}  & \multirow{3}{*}{\begin{tabular}[c]{@{}l@{}}WNS : 0.12ns\\ DRC Vio: 303\\ Density: 0.615\end{tabular}}   & \multirow{3}{*}{\begin{tabular}[c]{@{}l@{}}WNS : 0.119ns\\ DRC Vio: 0\\ Density: 0.615\end{tabular}}   \\
                                     &                                                                                                   &                                                                                  &                             &                                                                                                        &                                                                                                        &                                                                                                       &                                                                                                         &                                                                                                        \\
                                     &                                                                                                   &                                                                                  &                             &                                                                                                        &                                                                                                        &                                                                                                       &                                                                                                         &                                                                                                        \\
\multirow{3}{*}{Ethernet MAC}        & \multirow{3}{*}{\begin{tabular}[c]{@{}l@{}}10ns, WNS: 5.621ns\\ Setup successfully\end{tabular}}  & \multirow{3}{*}{\begin{tabular}[c]{@{}l@{}}Util: 0.60\\ DRC Vio: 0\end{tabular}} & \multirow{3}{*}{DRC Vio: 0} & \multirow{3}{*}{\begin{tabular}[c]{@{}l@{}}WNS : 5.62ns\\ DRC Vio: 25\\ Density: 0.62\end{tabular}}    & \multirow{3}{*}{\begin{tabular}[c]{@{}l@{}}WNS : 5.475ns\\ DRC Vio: 23\\ Density: 0.624\end{tabular}}  & \multirow{3}{*}{\begin{tabular}[c]{@{}l@{}}WNS : 5.470ns\\ DRC Vio: 0\\ Density: 0.628\end{tabular}}  & \multirow{3}{*}{\begin{tabular}[c]{@{}l@{}}WNS : 5.53ns\\ DRC Vio: 69\\ Density: 0.624\end{tabular}}    & \multirow{3}{*}{\begin{tabular}[c]{@{}l@{}}WNS : 5.53ns\\ DRC Vio: 0\\ Density: 0.625\end{tabular}}    \\
                                     &                                                                                                   &                                                                                  &                             &                                                                                                        &                                                                                                        &                                                                                                       &                                                                                                         &                                                                                                        \\
                                     &                                                                                                   &                                                                                  &                             &                                                                                                        &                                                                                                        &                                                                                                       &                                                                                                         &                                                                                                        \\ \hline
\end{tabular}

}
\end{table*}

%% file: conclusion.tex
\section{Conclusion}
In this work, we introduce \textsc{PDAgent-Bench}, the first comprehensive benchmark for systematic evaluation of LLM/VLMs agents in VLSI physical design, comprising 353 tasks across five capability axes and a unified multi-agent framework for closed-loop interaction with EDA tools.
Our evaluation of 11 frontier LLMs/VLMs reveals a substantial gap between conceptual understanding and tool-grounded execution. We anticipate \textsc{PDAgent-Bench} will accelerate the development of domain-specific LLM/VLMs agents for this underexplored but high-impact domain.

%% file: appendix.tex
\newpage

\appendix

\section{Appendix}

\subsection{Characteristics of PDAGENT-BENCH}

\begin{figure}[]
\centering
\includegraphics[width=1\linewidth]{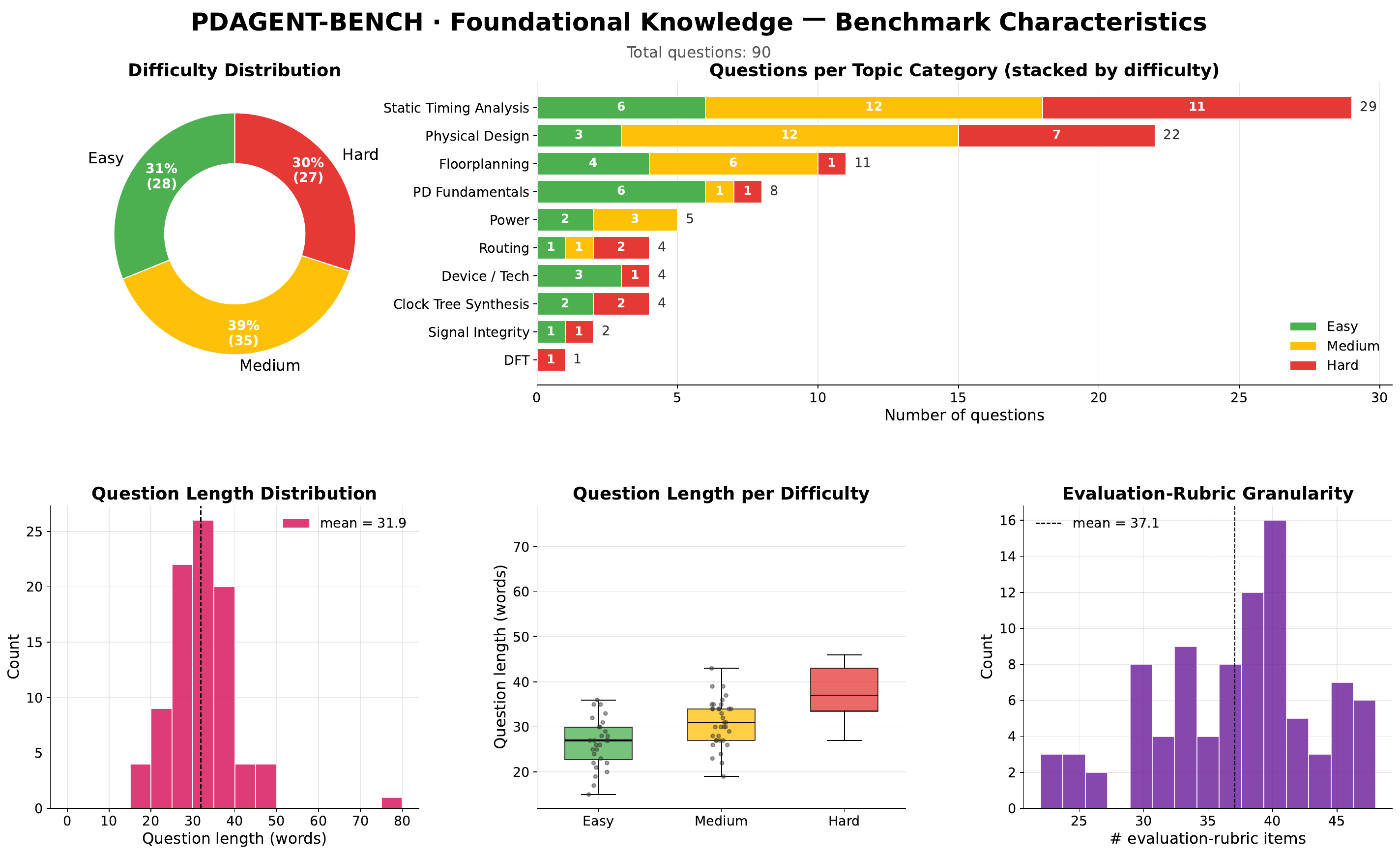}
\caption{Characteristics of the foundational-knowledge subset of \textsc{PDAGENT-BENCH} (90 questions). Questions are balanced across three difficulty levels and span ten physical-design topics, with timing analysis and physical design as the most represented categories. Each question is paired with a fine-grained evaluation rubric (mean of 37.1 items).}
\label{fig:basic_charac}
\end{figure}

\begin{figure*}[!t]
\centering
\includegraphics[width=1\linewidth]{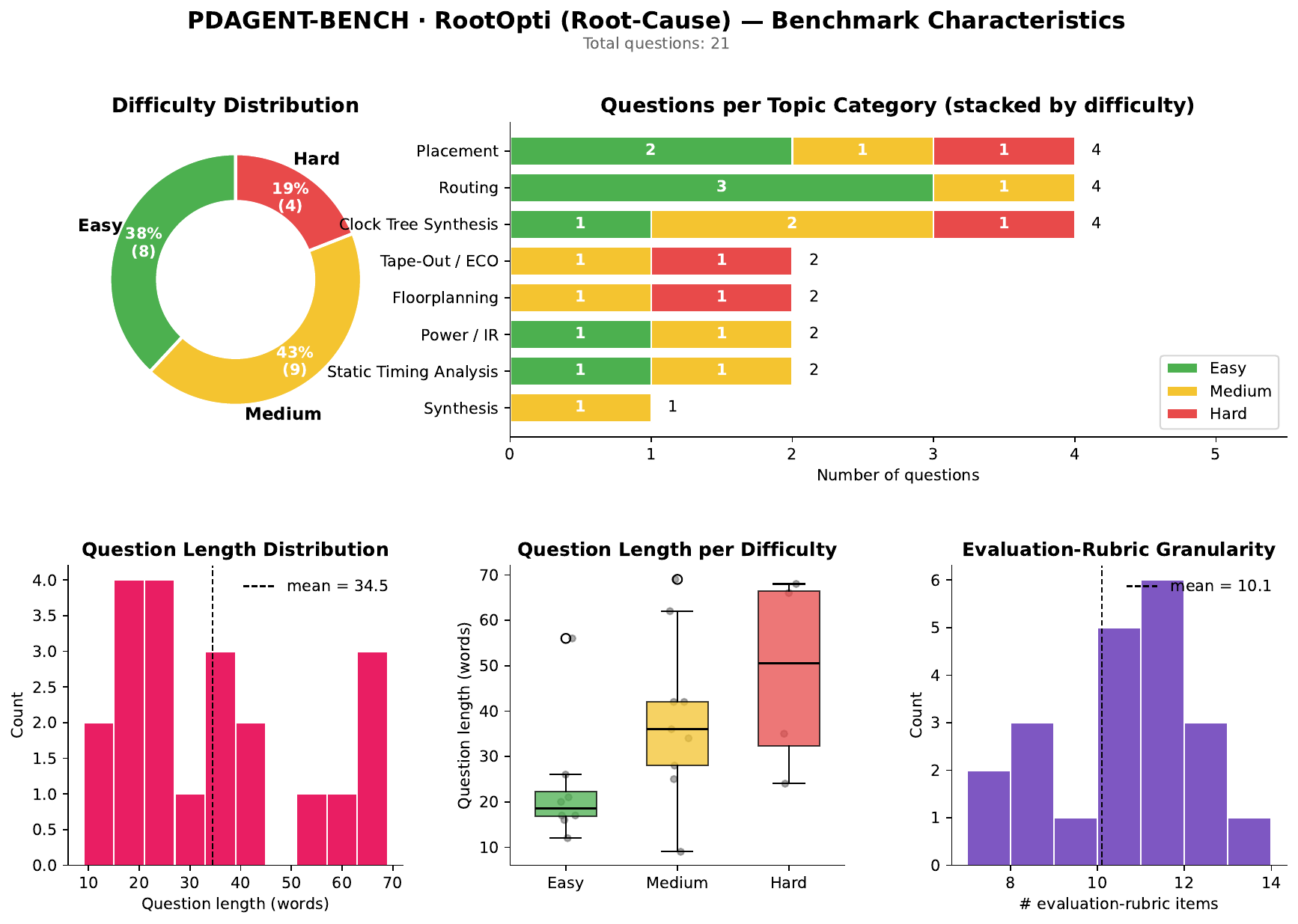}
\caption{Characteristics of the root-cause analysis subset of \textsc{PDAGENT-BENCH} (21 questions). Questions span eight physical-design stages, from synthesis through tape-out/ECO, with placement, routing, and clock tree synthesis as the most represented categories. Each question is paired with a targeted evaluation rubric (mean of 10.1 items).}
\label{fig:root_charac}
\end{figure*}

\begin{figure*}[!t]
\centering
\includegraphics[width=0.95\linewidth]{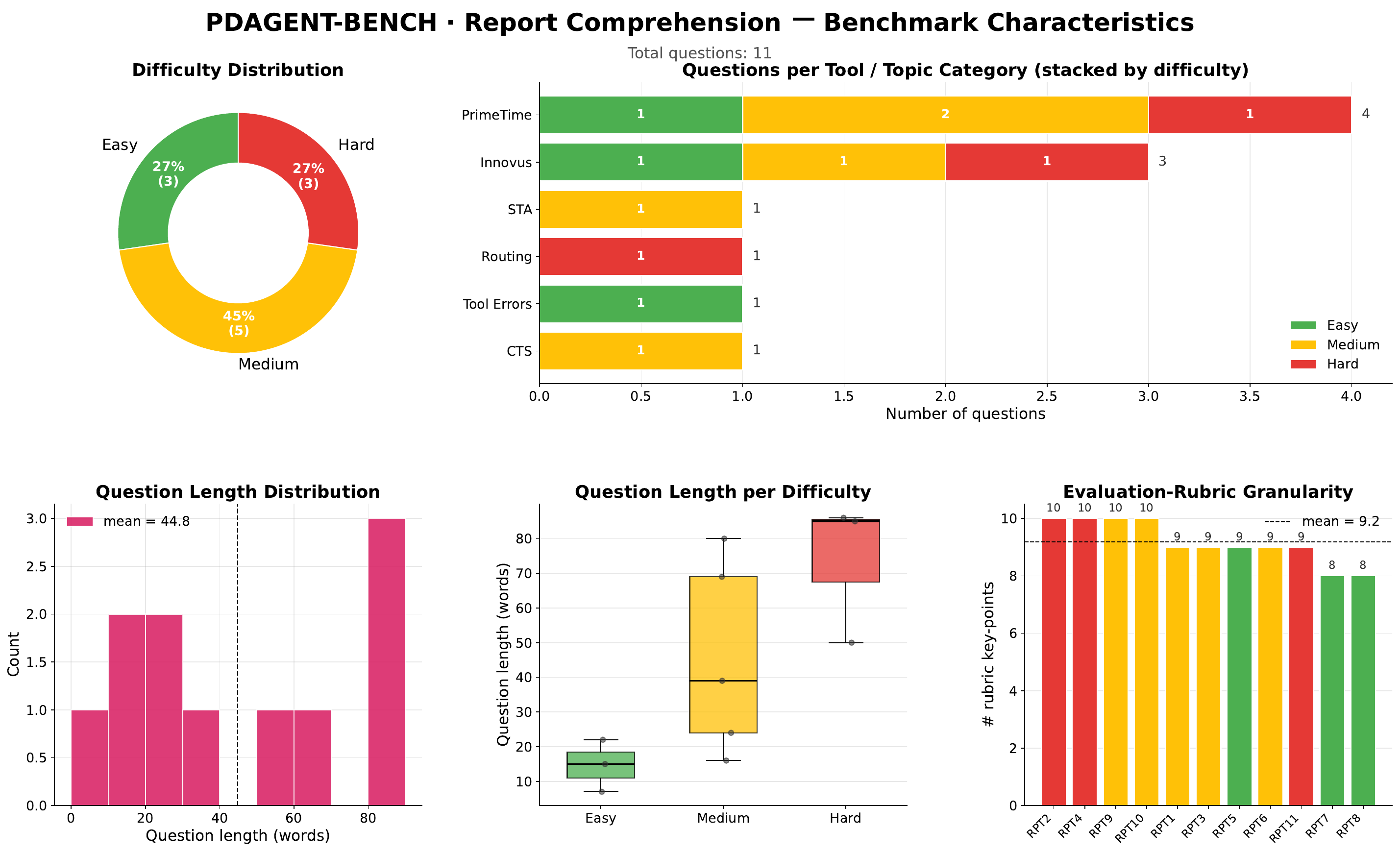}
\caption{Characteristics of the report-comprehension subset of \textsc{PDAgent-Bench} (11 questions). Questions cover reports from major EDA tools (PrimeTime, Innovus) across stages including STA, routing, CTS, and tool error logs. Each report is paired with a structured evaluation rubric.}
\label{fig:report_charac}
\end{figure*}

\begin{figure*}[!t]
\centering
\includegraphics[width=1\linewidth]{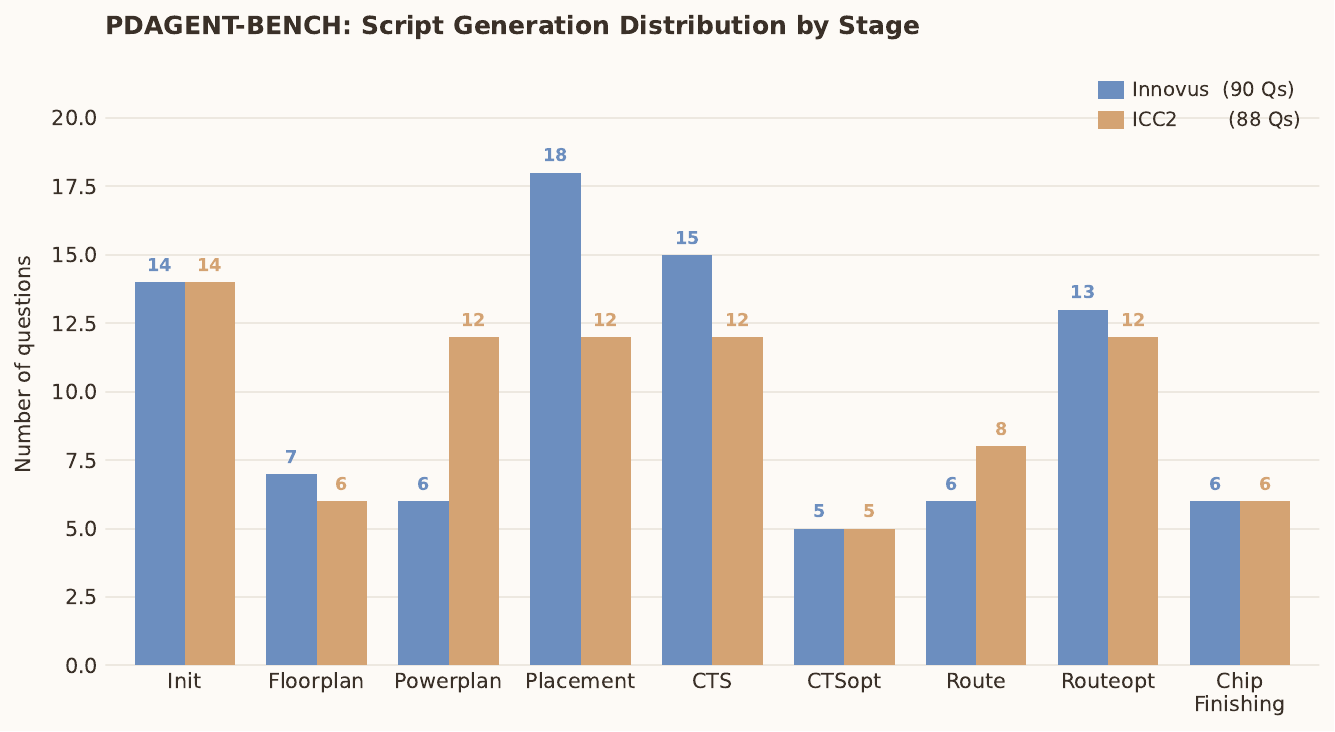}
\caption{Characteristics of the script generation subset of \textsc{PDAgent-Bench} (178 questions in total). Questions span nine physical-design stages, from initialization through chip finishing, and are provided in parallel for two industrial EDA tools, Innovus (90 questions) and ICC2 (88 questions). Placement, CTS, and routing-optimization stages account for the largest share of questions in both tools.}
\label{fig:script_charac}
\end{figure*}

We summarize the key characteristics of each subset in Figures~\ref{fig:basic_charac}--\ref{fig:script_charac}, including question count, difficulty distribution, topic coverage, and evaluation-rubric granularity. Together, the four subsets comprise 343 questions spanning the full physical-design workflow, from foundational concepts to tool-specific script generation.

\subsection{Error Case Analysis}
\label{app:error_case_analysis}

To understand the failure modes of current LLMs on physical-design tasks, we manually inspect the incorrect outputs of all evaluated models, categorize them by error type, and report the resulting distribution. The analysis surfaces the systematic weaknesses underlying the quantitative gaps reported in Table~\ref{tab:bench-models} and informs the directions discussed in Section~\ref{sec:findings}.

\begin{table}[ht]
\centering
\caption{Failure breakdown of ChatGPT-5.5 on \texttt{benchmark\_innovus.json} (52 failures out of 90).}
\label{tab:failure-breakdown}
\scalebox{0.85}{
\begin{tabular}{lcl}
\hline
\textbf{Category} & \textbf{Count} & \textbf{Description} \\
\hline
Multi-line near-miss              & 33 & Right intent/commands, wrong flag names or missing/extra steps \\
Same command, wrong attribute     & 12 & Single-line command correct, but flag is wrong \\
Same command, value formatting    & 3  & Right command + flags, formatting/value diff \\
Wrong command                     & 4  & Picked wrong command (incl.\ 1 free-text answer) \\
\hline
\textbf{Total}                    & \textbf{52} & \\
\hline
\end{tabular} }
\end{table}

\begin{table}[ht]
\centering
\caption{Examples of attribute-name hallucinations: the model picks the correct command but invents plausible-but-wrong Innovus flag names.}
\label{tab:attr-hallucinations}
\scalebox{0.92}{
\begin{tabular}{lll}
\hline
\textbf{ID} & \textbf{Expected flag} & \textbf{Generated flag} \\
\hline
TC\_016 & \texttt{-around core -center 1}             & \texttt{-follow core} \\
TC\_018 & \texttt{-subclass}                          & \texttt{-userDefined} \\
TC\_021 & \texttt{-no\_routing\_blkg}                 & \texttt{-ignore\_routing\_blockage} \\
TC\_026 & \texttt{-rule \ldots\ -wellCutCell}         & \texttt{-distance \ldots\ -boundaryCell} \\
TC\_038 & \texttt{-place\_opt\_save\_db}              & \texttt{-place\_global\_save\_db} \\
TC\_054 & \texttt{-outDir}                            & \texttt{-report\_dir} \\
TC\_069 & \texttt{-drouteMinSlackForWireOptimization} & \texttt{-droutePostRouteSpreadWireMinSlack} \\
TC\_077 & \texttt{lefDefOutVersion}                   & \texttt{defOutVersion} \\
TC\_084 & \texttt{specifyCellPad CLK*} (positional)   & \texttt{specifyCellPad -cell CLK*} \\
\hline
\end{tabular} }
\end{table}

\begin{table}[ht]
\centering
\caption{GPT-5.5 Basic Benchmark Executive Summary (90 questions).}
\label{tab:chatgpt55-basic-summary}
\renewcommand{\arraystretch}{1.25}
\scalebox{0.95}{
\begin{tabular}{@{} >{\bfseries\raggedright\arraybackslash}p{0.28\linewidth} p{0.66\linewidth} @{}}
\toprule
\multicolumn{2}{c}{\textbf{Overall Score: 724.0 / 900.5 = 80.4\%}} \\
\midrule
By Difficulty       & Easy 83.0\% \quad$\vert$\quad Medium 81.7\% \quad$\vert$\quad Hard 76.0\% \\
Catastrophic Fails ($<$60\%) & 2 / 90 \;($\rightarrow$ Q2: 15\%, \; Q86: 45\%) \\
Low Criteria ($\leq$50\% of max) & 32 / 454 ($\sim$7\% --- concentrated, not diffuse) \\
\midrule
\multicolumn{2}{c}{\textbf{Strengths vs.\ Weaknesses}} \\
\midrule
Strengths &
$\bullet$~File formats (GDSII / SDF / DEF, all $\geq$85\%) \newline
$\bullet$~STA definitions (slack, false path, best/worst path: 100\%) \newline
$\bullet$~Technology-node comparison (Q63: 100\%) \newline
$\bullet$~Digital logic fundamentals \newline
$\bullet$~Physical Design Fundamentals topic family (89\%) \\
Weaknesses &
$\bullet$~Question-interpretation errors (Q2) \newline
$\bullet$~Quantitative reasoning / layer math (Q86) \newline
$\bullet$~Library models --- NLDM vs.\ CCS (65\%) \newline
$\bullet$~CTS insertion-delay tuning \newline
$\bullet$~Post-route to signoff correlation \\
\midrule
\multicolumn{2}{c}{\textbf{Top 3 Error Patterns}} \\
\midrule
1 & Misreading the question's framing $\rightarrow$ wrong taxonomy (Q2). \\
2 & Quantitative reasoning with constraint propagation (Q86). \\
3 & ``Missing one column'' --- partial trade-off tables, lost bonus points. \\
\midrule
\multicolumn{2}{c}{\textbf{Verdict}} \\
\midrule
\multicolumn{2}{p{\linewidth}}{Strong general PD/STA knowledge; points are lost to under-elaboration and one-off interpretation errors, not deep knowledge gaps.} \\
\bottomrule
\end{tabular} }
\end{table}

\subsection{Results on Full Flow Implementation}
\label{app:whole_result_fullflow}

\input{tab_tsmc28nm}

\input{tab_standalone_full-flow}

The evaluation results of \textbf{PDAgent} on the PDAgent-Bench holistic flow suite are summarized in Table~\ref{tab:fullflow_results}, using GPT-5mini as the backbone model. The results demonstrate that \textbf{PDAgent} is capable of executing a holistic physical-design flow end-to-end, performing self-iterative design space exploration, autonomously resolving DRC violations, and optimizing the design toward timing closure. The consistent performance across diverse designs demonstrates the generality of our method. On average, \textbf{PDAgent} completes the full flow through 53 interactions with the EDA tools, reflecting the iterative nature of physical-design optimization.

Our method performs well on small and medium-complexity designs, achieving timing closure with zero or minimal DRC violations at the final RouteOpt stage. However, on larger and more complex designs such as NVDLA-large and Open910, residual DRC violations remain after RouteOpt, and timing closure is not fully achieved in all cases. We attribute these limitations not only to the inherent complexity of the design space at advanced technology nodes, but also to the current lack of more sophisticated ECO and DRC-fixing capabilities in our framework, which we identify as a promising direction for future work. We anticipate that future work building upon \textbf{PDAgent} can achieve improved timing slack, higher utilization, cleaner DRC results, and reduced interaction overhead through more advanced optimization strategies.

\textbf{Comparison Against RAG-enhanced Agent Baseline}

We further evaluate two RAG-enhanced LLM baselines, which retrieve relevant content from EDA command reference manuals and physical-design guidebooks, using GPT-5mini backbone models for fair comparison.The Full flow suite results are reported in Table~\ref{tab:rag_results}.

 For the Full flow suite, the RAG-enhanced method fails to complete most designs beyond the first two stages. We attribute this performance gap to the highly structured and interdependent nature of physical-design flows: each stage comprises multiple steps with precise tool-level syntax and context-specific requirements, which general-purpose retrieved documentation is too coarse-grained to fully capture. Unlike \textbf{PDAgent}, which encodes actionable, stage-specific skills, the RAG-based approach lacks the procedural knowledge necessary to handle tool-specific errors, adapt parameters across designs, and drive the flow to completion.

\subsection{Token Usage of full flow implementation}

\begin{table}[t]
\centering
\caption{Token usage of the full-flow implementation across all benchmark designs.
We report input tokens, output tokens, the resulting total, and the number of LLM calls.}
\label{tab:token-usage}
\footnotesize
\setlength{\tabcolsep}{6pt}
\begin{tabular}{l r r r r}
\toprule
Design & Input & Output & Total & LLM calls \\
\midrule
TinyRISCV           & 327{,}648    & 6{,}013  & 333{,}661    & 67  \\
Open910             & 1{,}283{,}641 & 24{,}167 & 1{,}307{,}808 & 182 \\
Systolic array      & 227{,}648    & 4{,}013  & 231{,}661    & 47  \\
16b pipelined mult. & 144{,}057    & 3{,}924  & 147{,}981    & 29  \\
AES-256             & 259{,}783    & 3{,}948  & 263{,}731    & 57  \\
NVDLA-small         & 1{,}259{,}783 & 21{,}948 & 1{,}281{,}731 & 137 \\
NVDLA-large         & 1{,}595{,}980 & 25{,}943 & 1{,}621{,}923 & 192 \\
UART controller     & 127{,}648    & 4{,}013  & 131{,}661    & 27  \\
Elevator FSM        & 174{,}057    & 4{,}524  & 178{,}581    & 34  \\
Ethernet MAC        & 444{,}057    & 8{,}924  & 452{,}981    & 79  \\

\bottomrule
\end{tabular}
\end{table}

Table~\ref{tab:token-usage} reports token consumption for the full flow across the
ten benchmarks. The cost is overwhelmingly dominated by
input (context) tokens: 5.84M. This is expected for an agentic
EDA loop, in which each step ingests large synthesis logs, timing reports, and
netlist summaries but emits only short commands and decisions. Token usage tracks
design complexity: NVDLA-large, the largest design ($1.48$M cells), is the most
expensive at $1.62$M tokens and $192$ calls, whereas the small control designs (UART,
Elevator FSM, 16b multiplier) converge in $27$--$34$ calls and under $180$k tokens
each.

\subsection{Macro Placement results}
\label{sec: macroplacement}

We evaluate on five ChiPBench designs against seven learning-based and analytical
placers (WireMask-EA, SA, MaskPlace, ChiPFormer, DREAMPlace, AutoDMP, and OpenROAD) to show that our platform can handle the complex macro placement. For each design, the
model generates five candidate placements that are refined in parallel, with each
candidate given up to five rounds of feedback-driven refinement.
As shown in Table~\ref{tab:macro_all}, our flow achieves the best WNS and TNS on all
four multi-baseline designs, cutting TNS by up to ${\sim}60\times$ over the strongest
baseline (e.g., $-244.94\!\rightarrow\!-4.01$ on ariane133) and staying on par on
swerv\_wrapper. It also attains the lowest power, at the
cost of larger area. On VeriGPU, where OpenROAD is the only baseline, it reaches a
comparable point with better TNS. Overall, an 8B vision-language backbone driven by
our flow can match or surpass specialized macro placers on timing closure.

\begin{table}[t]
\centering
\caption{Macro placement results across all benchmark designs (our method as \textit{Ours}).}
\label{tab:macro_all}
\small
\setlength{\tabcolsep}{5pt}
\begin{tabular}{l l rrrr}
\toprule
Design & Method & Power\,$\downarrow$ & WNS\,$\uparrow$ & TNS\,$\uparrow$ & Area\,$\downarrow$ \\
\midrule
\multirow{8}{*}{ariane133}
 & WireMask-EA & 0.369 & -0.417 & -329.353 & 349228 \\
 & SA          & 0.365 & -0.512 & -650.399 & 347043 \\
 & MaskPlace   & 0.373 & -0.349 & -244.936 & 357322 \\
 & ChiPFormer  & 0.370 & -0.553 & -860.952 & 349005 \\
 & DREAMPlace  & 0.367 & -0.540 & -690.266 & 348180 \\
 & AutoDMP     & 0.350 & -0.982 & -1913.660 & \textbf{344091} \\
 & OpenROAD    & 0.359 & -0.498 & -596.718 & 352706 \\
 & Ours        & \textbf{0.257} & \textbf{-0.08} & \textbf{-4.01} & 357917 \\
\midrule
\multirow{8}{*}{bp\_be}
 & WireMask-EA & 0.467 & -2.142 & -4093.970 & 123966 \\
 & SA          & 0.459 & -2.494 & -5510.140 & 124667 \\
 & MaskPlace   & 0.456 & -2.439 & -4459.870 & 124938 \\
 & ChiPFormer  & 0.425 & -2.169 & -3541.820 & \textbf{110904} \\
 & DREAMPlace  & 0.458 & -2.163 & -3648.020 & 125221 \\
 & AutoDMP     & 0.453 & -1.948 & -3351.120 & 125471 \\
 & OpenROAD    & 0.409 & -1.862 & -2343.290 & 118786 \\
 & Ours        & \textbf{0.421} & \textbf{-0.63} & \textbf{-57.67} & 128865 \\
\midrule
\multirow{8}{*}{bp\_fe}
 & WireMask-EA & 0.309 & -1.666 & -777.420 & 77648 \\
 & SA          & 0.310 & -1.179 & -1236.880 & 81919 \\
 & MaskPlace   & 0.309 & -1.318 & -771.363 & 77881 \\
 & ChiPFormer  & 0.299 & -1.197 & -1000.190 & \textbf{72723} \\
 & DREAMPlace  & 0.294 & -1.117 & -473.261 & 77427 \\
 & AutoDMP     & 0.312 & -1.253 & -1198.430 & 81294 \\
 & OpenROAD    & 0.285 & -1.092 & -491.698 & 75884 \\
 & Ours        & \textbf{0.24} & \textbf{-0.43} & \textbf{-29.06} & 81323 \\
\midrule
\multirow{8}{*}{swerv\_wrapper}
 & WireMask-EA & 0.666 & -1.027 & -873.506 & 205627 \\
 & SA          & 0.648 & -0.962 & -854.737 & \textbf{200280} \\
 & MaskPlace   & 0.690 & -1.251 & -1306.460 & 206481 \\
 & ChiPFormer  & 0.674 & -1.189 & -1282.240 & 205775 \\
 & DREAMPlace  & 0.646 & -1.061 & -780.196 & 203896 \\
 & AutoDMP     & 0.648 & -1.094 & -773.339 & 200823 \\
 & OpenROAD    & 0.628 & -1.127 & -1163.820 & 202853 \\
 & Ours        & \textbf{0.626} & \textbf{-0.67} & \textbf{-750.37} & 204713 \\
\midrule
\multirow{2}{*}{VeriGPU}
 & OpenROAD & \textbf{0.0779} & \textbf{-0.26} & -16.84 & \textbf{271391} \\
 & Ours     & 0.0931 & -0.28 & \textbf{-14.14} & 273414 \\
\bottomrule
\end{tabular}
\end{table}

\subsection{PDAgent Design Details}
\label{appendix:agent_details}

\subsubsection{Agent Design}
\label{sec:agent_design}

Each agent inherits from a common \texttt{BaseAgent} class that 
provides message handling, LLM access, tool query primitives, and a 
shared \texttt{DesignContext}. The \texttt{DesignContext} is a cached 
representation of the active design, populated by querying the EDA 
tool after initialization, including macro locations, die/core area, 
and clock topology. It is shared across all agents to maintain 
consistent design state throughout the optimization loop.

\noindent\textbf{Worker Agent.}
The Worker Agent maintains a persistent EDA session and executes 
subtasks in three modes: (1) skill-guided generation, where the LLM 
synthesizes Tcl scripts from skill content and design context; 
(2) script execution, where reference Tcl scripts are sourced and run 
directly; and (3) optimization application, where targeted 
optimizations requested by the Optimizer Agent are applied and 
coordinated through the Planner Agent. Executed scripts, logs, and 
layout snapshots are organized by stage in a session history with 
manifest tracking, ensuring full traceability and enabling downstream 
agents to reference artifacts from any prior stage.

\noindent\textbf{Analyzer Agent.}
The Analyzer Agent is selectively activated at analysis-critical stages 
(e.g., placement, CTS, and routing), where raw EDA tool logs are 
verbose and difficult to interpret directly. At each activated stage, 
it extracts a structured \texttt{StageMetrics} vector covering timing, 
utilization, routing, clock, and power metrics, and passes this 
condensed summary to the Debugger and Optimizer Agents for closed-loop 
refinement. Lightweight stages such as floorplanning and powerplanning 
bypass analysis, reducing runtime overhead without sacrificing result 
quality at critical design junctures.

\noindent\textbf{Debugger Agent.}
After the Worker Agent completes a stage, the Debugger Agent checks 
for DRC violations and other errors by executing an iterative 
diagnosis loop. It loads the stage-specific debug skill, initializes 
a multi-turn LLM conversation with the skill content, EDA output, and 
convergence state as context, then iteratively performs 
check$\rightarrow$fix$\rightarrow$check cycles via direct tool queries. Each step is 
tracked in an \texttt{AgentStageContext} that maintains the full 
conversation history, violations found and fixed, and budget 
consumption. Skills provide the check criteria and corresponding fix 
strategies for each stage.

To prevent unbounded iteration, a three-tier budget hierarchy is 
enforced: (1) micro-level, handling syntax errors with up to 2 
retries; (2) meso-level, handling parameter adjustments with up to 
4 retries; and (3) macro-level, handling cross-stage rollback with 
up to 2 retries. When the budget is exhausted, the Debugger Agent 
escalates to the Optimizer Agent for more aggressive intervention.

\noindent\textbf{Optimizer Agent.}
The Optimizer Agent operates in two modes. In iterative mode, it 
conducts a multi-turn LLM-guided loop analogous to the Debugger Agent: 
checking current metrics, adjusting parameters (e.g., placement 
effort illustrated in Figure~\ref{fig:openlayout_agent}(c)), informed by stage-specific optimization 
skills and the full iteration trajectory stored in \texttt{MetricsDB}, 
then verifying improvements. In tree-of-thought mode, triggered on 
escalation or LLM-directed exploration, it performs branching design 
space exploration by generating multiple solution candidates and 
executing them either in parallel or sequentially, as described in 
\S\ref{sec:dso}.

\subsubsection{Closed-Loop PPA Optimization Mechanism}
\label{sec:dso}

The PPA Optimization Mechanism in PDAgent operates at two levels: a 
\textbf{stage-level} that progresses through the physical design 
flow step by step, and a \textbf{parameter-level} that explores 
the design space within each stage.

\noindent\textbf{Stage-Level Execution.}
The Planning Agent maintains a plan $\mathcal{P} = \{s_1, \ldots, 
s_n\}$ of subtasks ordered by dependency. For each subtask $s_i$, 
execution follows one of two paths based on the tool output:

(1) \textbf{Success path:} Tool $\xrightarrow{\text{result}}$ Planner 
$\xrightarrow{\text{if worthy}}$ Analyzer $\xrightarrow{\text{metrics}}$ 
Optimizer $\xrightarrow{\text{decision}}$ Planner $\rightarrow$ 
advance or re-run;

(2) \textbf{Failure path:} Tool $\xrightarrow{\text{error}}$ Planner 
$\rightarrow$ Debugger $\xrightarrow{\text{action}}$ Planner 
$\rightarrow$ \{retry, fix, rollback, skip, escalate\}.
The convergence state, classified as improving, stalled, or diverging 
based on the metric trajectory across iterations, informs action 
selection at each decision point.

\noindent\textbf{Parameter-Level Exploration.}
When the Optimizer Agent determines that local parameter adjustments 
are insufficient, for example upon escalation from the Debugger Agent 
or detection of a stalled convergence state, it activates a 
tree-of-thought search over the parameter space. The search proceeds 
in batches: at each batch $b \in \{1, \ldots, B\}$, the LLM generates 
$K$ candidate parameter configurations $\{c_1^{(b)}, \ldots, 
c_K^{(b)}\}$, such as redoing macro placement or revising the power 
plan to reduce routing layer pressure, each accompanied by the 
corresponding EDA commands and evaluation queries. Each candidate is 
evaluated by saving a design checkpoint, applying the parameter 
modifications, executing the evaluation commands, and extracting the 
resulting metrics. The design is then restored to the checkpoint 
before the next candidate is evaluated.

The fitness of each candidate is computed as a stage-aware weighted 
score:
\begin{equation}
    \text{score}(c) = \sum_{m \in \mathcal{M}_s} w_m^{(s)} \cdot 
    \hat{v}_m(c),
\end{equation}
where $\mathcal{M}_s$ denotes the metric set for stage $s$, 
$w_m^{(s)}$ the stage-specific weight, and $\hat{v}_m(c) \in [0, 1]$ 
the normalized metric value. The best candidate $c^* = 
\arg\max_c \text{score}(c)$ becomes the starting point for batch 
$b{+}1$. The search terminates when stage requirements are satisfied 
(e.g., DRC count $= 0$ for routing), the score improvement falls 
below threshold $\delta$, or the batch budget $B$ is exhausted.

\noindent\textbf{Iteration Tracking.}
A persistent \texttt{MetricsDB} records the full parameter to metric 
mapping across all iterations, including parameter provenance, 
capturing which agent modified which parameter and the rationale 
behind each change. This trajectory data serves two purposes: it 
provides the Optimizer Agent with historical context for informed 
tuning decisions, and it enables post-hoc analysis of the DSO process 
for reproducibility and insight.

\subsection{System Prompts}
\label{system_prompt}

We list the system and judge prompts used throughout our evaluation in Figures~\ref{fig:question_prompt}--\ref{fig:score_prompt}. Figure~\ref{fig:question_prompt} shows the system prompt provided to all evaluated models on the Physical Design Understanding tasks (Basic, RootOpt, Report, STA). Figure~\ref{fig:script_prompt} shows the system prompt for the script-generation tasks (Innovus, ICC2, ECO, FM). Figure~\ref{fig:score_prompt} shows the prompt given to our LLM-Judge (Claude Sonnet 4.6) for automatic scoring of free-response answers.

\begin{figure}
    \begin{tcolorbox}
    You are a senior VLSI physical design engineer. Answer the user's question with technically accurate, well-structured explanations. Cover key concepts, trade-offs, and any relevant tool/flow details. Be concise, compact but complete.
    \end{tcolorbox}
    \caption{System prompt used for Physical Design Understanding tasks.}
    \label{fig:question_prompt}
\end{figure}

\begin{figure}[h]
    \begin{tcolorbox}
    You are a senior VLSI physical design engineer. Answer each question with ONLY the exact Tcl/shell command(s) required. Do not add explanations, comments, markdown fences, or surrounding prose.
    \end{tcolorbox}
    \caption{System prompt used at inference time for the script Generation tasks (Innovus, ICC2, ECO, FM).}
    \label{fig:script_prompt}
\end{figure}

\begin{figure}[h]
    \begin{tcolorbox}
    Please act as an impartial physical-design engineer and judge the quality of the response provided by an AI assistant to the user question shown below. Evaluate the response objectively against the rubric supplied with each data item.
    
    Begin your evaluation with a brief explanation justifying your assessment. Be as fair and objective as possible.
    
    After the explanation, conclude with a numerical score from 0 to 10 in the following strict format:
    
    \textbf{score: X}
    
    For example: \textbf{score: 7}
    \end{tcolorbox}
    \caption{LLM-Judge prompt used to score model responses on the Physical Design Understanding tasks.}
    \label{fig:score_prompt}
\end{figure}

\subsection{Examples from PDAgent-Bench}

We illustrate the structure and content of PDAgent-Bench through representative examples drawn from each task category. Figures~\ref{fig:example_question}--\ref{fig:example_rubric} show a sample question, reference answer, and scoring rubric from the Basic benchmark. The complete benchmark, along with all model outputs and scoring results, is included in the supplementary material.

\begin{tcolorbox}[breakable]
    \textbf{Question.} In VLSI backend physical design flow, describe the essential input files required by a physical design tool and the key output files it generates. For each file, briefly explain its purpose and common format extensions.
    
    \medskip
    \textbf{Reference Answer.}
    
    \textit{Input data required for physical design:}
    \begin{enumerate}[noitemsep, topsep=2pt, leftmargin=*]
      \item \textbf{Technology file} (\texttt{.tf} in Synopsys, \texttt{.techlef} in Cadence): describes units, drawing patterns, layers, design rules, vias, and parasitic R/C of the manufacturing process.
      \item \textbf{Physical libraries} (\texttt{.lef}, \texttt{.gds}; or Synopsys \texttt{.CEL}, \texttt{.FRAM}): layout information and abstract models for placement and routing (pin accessibility, blockages, etc.).
      \item \textbf{Timing, logical, and power libraries} (\texttt{.lib}, or LM view \texttt{.db}): timing and power information for all design elements.
      \item \textbf{TDF file} (\texttt{.tdf} / \texttt{.io}): pad/pin arrangement (order and location); for full-chip flows also captures VDD/VSS pads and power-cut diodes not present in the Verilog netlist.
      \item \textbf{Constraints} (\texttt{.sdc}): area, power, and timing constraints.
      \item \textbf{PDEF} (optional): row and cell placement locations.
      \item \textbf{DEF} (optional): row, cell, and pre-existing placement information.
    \end{enumerate}
    
    \textit{Output data from a physical design tool:}
    \begin{enumerate}[noitemsep, topsep=2pt, leftmargin=*]
      \item \textbf{Standard Delay Format} (\texttt{.sdf}): post-layout timing (excluding load information).
      \item \textbf{Parasitics} (\texttt{.spef}, \texttt{.dspf}): extracted R/C of cells and nets.
      \item \textbf{Post-routed netlist} (\texttt{.v}; flat or hierarchical): connectivity of all cells.
      \item \textbf{Physical layout} (\texttt{.gds}): final layout for tape-out.
      \item \textbf{Design Exchange Format} (\texttt{.def}): row, cell, and net placement locations.
    \end{enumerate}
\end{tcolorbox}
\captionof{figure}{Example question and reference answer from the Basic benchmark (Physical Design Understanding).}
\label{fig:example_question}

\bigskip

\begin{tcolorbox}[breakable]
    \textbf{Evaluation Rubric} (total: 10 points)
    \begin{enumerate}[noitemsep, topsep=2pt, leftmargin=*]
      \item \textbf{Input Files --- Technology \& Libraries (3 pts).} 1 pt each for: (a) technology file (\texttt{.tf} / \texttt{.techlef}) and its role in process rules and parasitics; (b) physical libraries (\texttt{.lef} / \texttt{.gds} / \texttt{.cel} / \texttt{.fram}) and their role in layout abstraction; (c) timing/power libraries (\texttt{.lib} / \texttt{.db}) and their role in timing/power characterization.
      \item \textbf{Input Files --- Constraints \& Optional Files (2 pts).} 1 pt each for: (a) SDC constraints file (timing, area, power); (b) IO/TDF or DEF/PDEF for pad placement or pre-existing placement data.
      \item \textbf{Output Files --- Timing \& Parasitics (2 pts).} 1 pt each for: (a) SDF file for post-layout timing delays; (b) SPEF or DSPF for extracted RC parasitics.
      \item \textbf{Output Files --- Netlist, Layout \& DEF (2 pts).} 1 pt for the post-routed Verilog netlist (\texttt{.v}, flat or hierarchical); 0.5 pt for GDS (physical layout); 0.5 pt for DEF (final placement and routing data).
      \item \textbf{Clarity, Completeness \& Technical Accuracy (1 pt).} Answer is well-organized (inputs vs.\ outputs clearly separated), file extensions correctly associated with descriptions, and no significant factual errors or omissions of major file types.
    \end{enumerate}
\end{tcolorbox}
\captionof{figure}{Scoring rubric corresponding to Figure~\ref{fig:example_question}, used by the LLM-Judge to grade model responses.}
\label{fig:example_rubric}

\subsection{Privacy-Preserving Deployment with Multiple Commercial Parties}
\label{sec: privacy-preserving}

The agentic PD design workflow involves collaboration among commercial EDA tools, PDKs, and LLMs/VLMs, all of which must remain proprietary.
To date, there are still no standards or policies to guide such design workflows.
This creates major deployment and operability barriers for the community, as highlighted at the recent first Cognitive Silicon Design Workshop (CSDW)~\cite{CSDW}.
To maximally preserve privacy between these parties, our framework uses a Python coordinator that runs on local hardware/environment and mediates all interactions between commercial EDA tools, PDKs, and foundation models. 
Rather than allowing an LLM/VLM to invoke tools directly, the coordinator executes each tool command and writes its output to predefined disk locations.
A deterministic parser then extracts only task-relevant fields, such as sign-off metrics (WNS, DRC count, and utilization) and compact error signatures. 
Consequently, the model is never exposed to raw tool dumps, foundry libraries, technology LEF/LIB files, or other proprietary artifacts.
It receives only a narrow, sanitizable slice of information explicitly selected by the coordinator. 
This design makes the privacy boundary an enforceable software interface under our control, rather than relying on the model to prevent the leakage or misuse of sensitive information.

%% file: tab_tsmc28nm.tex
\begin{table*}[t]
\centering
\caption{Full-flow implementation results of \textbf{PDAgent} on the 
PDAgent-Bench full flow suite (TSMC 28nm). WNS and DRC violations are reported 
at each stage; positive WNS indicates timing closure. Utilization (Util.) 
is reported after floorplanning.}
\label{tab:fullflow_results}

\scalebox{0.56}{
\begin{tabular}{lllllllll}
\hline
Design                               & init                                                                                              & floorplan                                                                        & powerplan                   & place                                                                                                  & CTS                                                                                                    & CTSopt                                                                                                & Route                                                                                                   & Routeopt                                                                                               \\ \hline
\multirow{3}{*}{TinyRISCV}           & \multirow{3}{*}{\begin{tabular}[c]{@{}l@{}}5ns, WNS: 1.362ns\\ Setup successfully\end{tabular}}   & \multirow{3}{*}{\begin{tabular}[c]{@{}l@{}}Util: 0.65\\ DRC Vio: 0\end{tabular}} & \multirow{3}{*}{DRC Vio: 0} & \multirow{3}{*}{\begin{tabular}[c]{@{}l@{}}WNS : -0.014ns\\ DRC Vio: 0\\ Density: 0.715\end{tabular}}  & \multirow{3}{*}{\begin{tabular}[c]{@{}l@{}}WNS : -0.127ns\\ DRC Vio: 0\\ Density: 0.7165\end{tabular}} & \multirow{3}{*}{\begin{tabular}[c]{@{}l@{}}WNS : 0.001ns\\ DRC Vio: 0\\ Density: 0.713\end{tabular}}  & \multirow{3}{*}{\begin{tabular}[c]{@{}l@{}}WNS : 0.021ns\\ DRC Vio: 1\\ Density: 0.722\end{tabular}}    & \multirow{3}{*}{\begin{tabular}[c]{@{}l@{}}WNS : 0.021ns\\ DRC Vio: 0\\ Density: 0.722\end{tabular}}   \\
                                     &                                                                                                   &                                                                                  &                             &                                                                                                        &                                                                                                        &                                                                                                       &                                                                                                         &                                                                                                        \\
                                     &                                                                                                   &                                                                                  &                             &                                                                                                        &                                                                                                        &                                                                                                       &                                                                                                         &                                                                                                        \\
\multirow{3}{*}{Open910}             & \multirow{3}{*}{\begin{tabular}[c]{@{}l@{}}1.5ns, WNS: 0.465ns\\ Setup successfully\end{tabular}} & \multirow{3}{*}{\begin{tabular}[c]{@{}l@{}}Util: 0.60\\ DRC Vio: 0\end{tabular}} & \multirow{3}{*}{DRC Vio: 0} & \multirow{3}{*}{\begin{tabular}[c]{@{}l@{}}WNS : -0.05ns\\ DRC Vio: 55\\ Density: 0.584\end{tabular}}  & \multirow{3}{*}{\begin{tabular}[c]{@{}l@{}}WNS : -0.009ns\\ DRC Vio: 2\\ Density: 0.653\end{tabular}}  & \multirow{3}{*}{\begin{tabular}[c]{@{}l@{}}WNS : 0.001ns\\ DRC Vio: 0\\ Density: 0.654\end{tabular}}  & \multirow{3}{*}{\begin{tabular}[c]{@{}l@{}}WNS : -0.431ns\\ DRC Vio: 127\\ Density: 0.703\end{tabular}} & \multirow{3}{*}{\begin{tabular}[c]{@{}l@{}}WNS : 0.008ns\\ DRC Vio: 125\\ Density: 0.731\end{tabular}} \\
                                     &                                                                                                   &                                                                                  &                             &                                                                                                        &                                                                                                        &                                                                                                       &                                                                                                         &                                                                                                        \\
                                     &                                                                                                   &                                                                                  &                             &                                                                                                        &                                                                                                        &                                                                                                       &                                                                                                         &                                                                                                        \\
\multirow{3}{*}{Systolic array}      & \multirow{3}{*}{\begin{tabular}[c]{@{}l@{}}3.4ns, WNS: 1.11ns\\ Setup successfully\end{tabular}}  & \multirow{3}{*}{\begin{tabular}[c]{@{}l@{}}Util: 0.5\\ DRC Vio: 0\end{tabular}}  & \multirow{3}{*}{DRC Vio: 0} & \multirow{3}{*}{\begin{tabular}[c]{@{}l@{}}WNS : 0.013ns\\ DRC Vio: 0\\ Density: 0.464\end{tabular}}   & \multirow{3}{*}{\begin{tabular}[c]{@{}l@{}}WNS : 0.003ns\\ DRC Vio: 0\\ Density: 0.464\end{tabular}}   & \multirow{3}{*}{\begin{tabular}[c]{@{}l@{}}WNS : 0.003ns\\ DRC Vio: 0\\ Density: 0.464\end{tabular}}  & \multirow{3}{*}{\begin{tabular}[c]{@{}l@{}}WNS : -0.312ns\\ DRC Vio: 0\\ Density: 0.464\end{tabular}}   & \multirow{3}{*}{\begin{tabular}[c]{@{}l@{}}WNS : 0.002ns\\ DRC Vio: 0\\ Density: 0.466\end{tabular}}   \\
                                     &                                                                                                   &                                                                                  &                             &                                                                                                        &                                                                                                        &                                                                                                       &                                                                                                         &                                                                                                        \\
                                     &                                                                                                   &                                                                                  &                             &                                                                                                        &                                                                                                        &                                                                                                       &                                                                                                         &                                                                                                        \\
\multirow{3}{*}{16b pipelined mult.} & \multirow{3}{*}{\begin{tabular}[c]{@{}l@{}}2.6ns, WNS: 0.808ns\\ Setup successfully\end{tabular}} & \multirow{3}{*}{\begin{tabular}[c]{@{}l@{}}Util: 0.55\\ DRC Vio: 0\end{tabular}} & \multirow{3}{*}{DRC Vio: 0} & \multirow{3}{*}{\begin{tabular}[c]{@{}l@{}}WNS : 0.02ns\\ DRC Vio: 0\\ Density: 0.625\end{tabular}}    & \multirow{3}{*}{\begin{tabular}[c]{@{}l@{}}WNS : -0.008ns\\ DRC Vio: 0\\ Density: 0.6289\end{tabular}} & \multirow{3}{*}{\begin{tabular}[c]{@{}l@{}}WNS : 0.012ns\\ DRC Vio: 0\\ Density: 0.629\end{tabular}}  & \multirow{3}{*}{\begin{tabular}[c]{@{}l@{}}WNS : 0.251ns\\ DRC Vio: 0\\ Density: 0.629\end{tabular}}    & \multirow{3}{*}{\begin{tabular}[c]{@{}l@{}}WNS : 0.251ns\\ DRC Vio: 0\\ Density: 0.629\end{tabular}}   \\
                                     &                                                                                                   &                                                                                  &                             &                                                                                                        &                                                                                                        &                                                                                                       &                                                                                                         &                                                                                                        \\
                                     &                                                                                                   &                                                                                  &                             &                                                                                                        &                                                                                                        &                                                                                                       &                                                                                                         &                                                                                                        \\
\multirow{3}{*}{AES-256}             & \multirow{3}{*}{\begin{tabular}[c]{@{}l@{}}3.5ns, WNS: 1.003ns\\ Setup successfully\end{tabular}} & \multirow{3}{*}{\begin{tabular}[c]{@{}l@{}}Util: 0.60\\ DRC Vio: 0\end{tabular}} & \multirow{3}{*}{DRC Vio: 0} & \multirow{3}{*}{\begin{tabular}[c]{@{}l@{}}WNS : 0.003ns\\ DRC Vio: 0\\ Density: 0.613\end{tabular}}   & \multirow{3}{*}{\begin{tabular}[c]{@{}l@{}}WNS : 0.006ns\\ DRC Vio: 0\\ Density: 0.615\end{tabular}}   & \multirow{3}{*}{\begin{tabular}[c]{@{}l@{}}WNS : 0.006ns\\ DRC Vio: 0\\ Density: 0.615\end{tabular}}  & \multirow{3}{*}{\begin{tabular}[c]{@{}l@{}}WNS : 0.12ns\\ DRC Vio: 303\\ Density: 0.615\end{tabular}}   & \multirow{3}{*}{\begin{tabular}[c]{@{}l@{}}WNS : 0.119ns\\ DRC Vio: 0\\ Density: 0.615\end{tabular}}   \\
                                     &                                                                                                   &                                                                                  &                             &                                                                                                        &                                                                                                        &                                                                                                       &                                                                                                         &                                                                                                        \\
                                     &                                                                                                   &                                                                                  &                             &                                                                                                        &                                                                                                        &                                                                                                       &                                                                                                         &                                                                                                        \\
\multirow{3}{*}{NVDLA-small}         & \multirow{3}{*}{\begin{tabular}[c]{@{}l@{}}2ns, WNS: 0.65ns\\ Setup successfully\end{tabular}}    & \multirow{3}{*}{\begin{tabular}[c]{@{}l@{}}Util: 0.60\\ DRC Vio: 0\end{tabular}} & \multirow{3}{*}{DRC Vio: 0} & \multirow{3}{*}{\begin{tabular}[c]{@{}l@{}}WNS : -1.06ns\\ DRC Vio: 46\\ Density: 0.424\end{tabular}}  & \multirow{3}{*}{\begin{tabular}[c]{@{}l@{}}WNS : 0.002ns\\ DRC Vio: 0\\ Density: 0.554\end{tabular}}   & \multirow{3}{*}{\begin{tabular}[c]{@{}l@{}}WNS : 0.002ns\\ DRC Vio: 0\\ Density: 0.554\end{tabular}}  & \multirow{3}{*}{\begin{tabular}[c]{@{}l@{}}WNS : -0.005ns\\ DRC Vio: 21\\ Density: 0.556\end{tabular}}  & \multirow{3}{*}{\begin{tabular}[c]{@{}l@{}}WNS : 0.002ns\\ DRC Vio: 19\\ Density: 0.556\end{tabular}}  \\
                                     &                                                                                                   &                                                                                  &                             &                                                                                                        &                                                                                                        &                                                                                                       &                                                                                                         &                                                                                                        \\
                                     &                                                                                                   &                                                                                  &                             &                                                                                                        &                                                                                                        &                                                                                                       &                                                                                                         &                                                                                                        \\
\multirow{3}{*}{NVDLA-large}         & \multirow{3}{*}{\begin{tabular}[c]{@{}l@{}}2.5ns, WNS: 0.69ns\\ Setup successfully\end{tabular}}  & \multirow{3}{*}{\begin{tabular}[c]{@{}l@{}}Util: 0.60\\ DRC Vio: 0\end{tabular}} & \multirow{3}{*}{DRC Vio: 0} & \multirow{3}{*}{\begin{tabular}[c]{@{}l@{}}WNS : -0.024ns\\ DRC Vio: 68\\ Density: 0.487\end{tabular}} & \multirow{3}{*}{\begin{tabular}[c]{@{}l@{}}WNS : -0.145ns\\ DRC Vio: 0\\ Density: 0.586\end{tabular}}  & \multirow{3}{*}{\begin{tabular}[c]{@{}l@{}}WNS : -0.063ns\\ DRC Vio: 0\\ Density: 0.594\end{tabular}} & \multirow{3}{*}{\begin{tabular}[c]{@{}l@{}}WNS : 0.021ns\\ DRC Vio: 242\\ Density: 0.596\end{tabular}}  & \multirow{3}{*}{\begin{tabular}[c]{@{}l@{}}WNS : 0.021ns\\ DRC Vio: 85\\ Density: 0.598\end{tabular}}  \\
                                     &                                                                                                   &                                                                                  &                             &                                                                                                        &                                                                                                        &                                                                                                       &                                                                                                         &                                                                                                        \\
                                     &                                                                                                   &                                                                                  &                             &                                                                                                        &                                                                                                        &                                                                                                       &                                                                                                         &                                                                                                        \\
\multirow{3}{*}{UART controller}     & \multirow{3}{*}{\begin{tabular}[c]{@{}l@{}}3.4ns, WNS: 1.108ns\\ Setup successfully\end{tabular}} & \multirow{3}{*}{\begin{tabular}[c]{@{}l@{}}Util: 0.65\\ DRC Vio: 0\end{tabular}} & \multirow{3}{*}{DRC Vio: 0} & \multirow{3}{*}{\begin{tabular}[c]{@{}l@{}}WNS : 1.1ns\\ DRC Vio: 0\\ Density: 0.782\end{tabular}}     & \multirow{3}{*}{\begin{tabular}[c]{@{}l@{}}WNS : 1.4ns\\ DRC Vio: 0\\ Density: 0.789\end{tabular}}     & \multirow{3}{*}{\begin{tabular}[c]{@{}l@{}}WNS : 1.59ns\\ DRC Vio: 0\\ Density: 0.789\end{tabular}}   & \multirow{3}{*}{\begin{tabular}[c]{@{}l@{}}WNS : -0.445ns\\ DRC Vio: 0\\ Density: 0.85\end{tabular}}    & \multirow{3}{*}{\begin{tabular}[c]{@{}l@{}}WNS : -0.09ns\\ DRC Vio: 0\\ Density: 0.85\end{tabular}}    \\
                                     &                                                                                                   &                                                                                  &                             &                                                                                                        &                                                                                                        &                                                                                                       &                                                                                                         &                                                                                                        \\
                                     &                                                                                                   &                                                                                  &                             &                                                                                                        &                                                                                                        &                                                                                                       &                                                                                                         &                                                                                                        \\
\multirow{3}{*}{Elevator FSM}        & \multirow{3}{*}{\begin{tabular}[c]{@{}l@{}}2.5ns, WNS: 0.708ns\\ Setup successfully\end{tabular}} & \multirow{3}{*}{\begin{tabular}[c]{@{}l@{}}Util: 0.7\\ DRC Vio: 0\end{tabular}}  & \multirow{3}{*}{DRC Vio: 0} & \multirow{3}{*}{\begin{tabular}[c]{@{}l@{}}WNS : 0.703ns\\ DRC Vio: 2\\ Density: 0.664\end{tabular}}   & \multirow{3}{*}{\begin{tabular}[c]{@{}l@{}}WNS : 0.702ns\\ DRC Vio: 2\\ Density: 0.665\end{tabular}}   & \multirow{3}{*}{\begin{tabular}[c]{@{}l@{}}WNS : 0.703ns\\ DRC Vio: 0\\ Density: 0.667\end{tabular}}  & \multirow{3}{*}{\begin{tabular}[c]{@{}l@{}}WNS : 0.622ns\\ DRC Vio: 0\\ Density: 0.667\end{tabular}}    & \multirow{3}{*}{\begin{tabular}[c]{@{}l@{}}WNS : 0.622ns\\ DRC Vio: 0\\ Density: 0.667\end{tabular}}   \\
                                     &                                                                                                   &                                                                                  &                             &                                                                                                        &                                                                                                        &                                                                                                       &                                                                                                         &                                                                                                        \\
                                     &                                                                                                   &                                                                                  &                             &                                                                                                        &                                                                                                        &                                                                                                       &                                                                                                         &                                                                                                        \\
\multirow{3}{*}{Ethernet MAC}        & \multirow{3}{*}{\begin{tabular}[c]{@{}l@{}}10ns, WNS: 5.621ns\\ Setup successfully\end{tabular}}  & \multirow{3}{*}{\begin{tabular}[c]{@{}l@{}}Util: 0.60\\ DRC Vio: 0\end{tabular}} & \multirow{3}{*}{DRC Vio: 0} & \multirow{3}{*}{\begin{tabular}[c]{@{}l@{}}WNS : 5.62ns\\ DRC Vio: 25\\ Density: 0.62\end{tabular}}    & \multirow{3}{*}{\begin{tabular}[c]{@{}l@{}}WNS : 5.475ns\\ DRC Vio: 23\\ Density: 0.624\end{tabular}}  & \multirow{3}{*}{\begin{tabular}[c]{@{}l@{}}WNS : 5.470ns\\ DRC Vio: 0\\ Density: 0.628\end{tabular}}  & \multirow{3}{*}{\begin{tabular}[c]{@{}l@{}}WNS : 5.53ns\\ DRC Vio: 69\\ Density: 0.624\end{tabular}}    & \multirow{3}{*}{\begin{tabular}[c]{@{}l@{}}WNS : 5.53ns\\ DRC Vio: 0\\ Density: 0.625\end{tabular}}    \\
                                     &                                                                                                   &                                                                                  &                             &                                                                                                        &                                                                                                        &                                                                                                       &                                                                                                         &                                                                                                        \\
                                     &                                                                                                   &                                                                                  &                             &                                                                                                        &                                                                                                        &                                                                                                       &                                                                                                         &                                                                                                        \\ \hline
\end{tabular}

}

\end{table*}

%% file: tab_standalone_full-flow.tex
\begin{table}[]
\centering
\caption{Early-stage physical design results using a RAG-based method without skills 
on the PDAgent-Bench full flow suite (TSMC 28nm). Results cover the first 
three implementation stages: synthesis, floorplanning, and power planning. 
The RAG approach exhibits consistent failures across all stages, 
including setup errors, pin placement failures, and incomplete 
power networks, demonstrating its limited capability in automated physical 
design.}
\label{tab:rag_results}

\scalebox{0.80}{
\begin{tabular}{llll}
\hline
Design                               & init                                                                                              & floorplan                                                                        & powerplan                                                                                          \\ \hline
\multirow{3}{*}{TinyRISCV}           & \multirow{3}{*}{\begin{tabular}[c]{@{}l@{}}5ns, WNS: 1.362ns\\ Setup failed\end{tabular}}         & \multirow{3}{*}{\begin{tabular}[c]{@{}l@{}}Pin placement \\ failed\end{tabular}} & \multirow{3}{*}{\begin{tabular}[c]{@{}l@{}}Failed to use \\ the correct \\ layer\end{tabular}}     \\
                                     &                                                                                                   &                                                                                  &                                                                                                    \\
                                     &                                                                                                   &                                                                                  &                                                                                                    \\
\multirow{3}{*}{Open910}             & \multirow{3}{*}{\begin{tabular}[c]{@{}l@{}}1.5ns, WNS: 0.465ns\\ Setup failed\end{tabular}}       & \multirow{3}{*}{\begin{tabular}[c]{@{}l@{}}Pin placement \\ failed\end{tabular}} & \multirow{3}{*}{\begin{tabular}[c]{@{}l@{}}Failed to \\ addRing\end{tabular}}                      \\
                                     &                                                                                                   &                                                                                  &                                                                                                    \\
                                     &                                                                                                   &                                                                                  &                                                                                                    \\
\multirow{3}{*}{Systolic array}      & \multirow{3}{*}{\begin{tabular}[c]{@{}l@{}}3.4ns, WNS: 1.11ns\\ Setup successfully\end{tabular}}  & \multirow{3}{*}{\begin{tabular}[c]{@{}l@{}}Pin placement \\ failed\end{tabular}} & \multirow{3}{*}{DRC Vio: 346}                                                                      \\
                                     &                                                                                                   &                                                                                  &                                                                                                    \\
                                     &                                                                                                   &                                                                                  &                                                                                                    \\
\multirow{3}{*}{16b pipelined mult.} & \multirow{3}{*}{\begin{tabular}[c]{@{}l@{}}2.6ns, WNS: 0.808ns\\ Setup failed\end{tabular}}       & \multirow{3}{*}{\begin{tabular}[c]{@{}l@{}}Util: 0.65\\ DRC Vio: 0\end{tabular}} & \multirow{3}{*}{\begin{tabular}[c]{@{}l@{}}Missing Bottom \\ and Top Power \\ Rings\end{tabular}}  \\
                                     &                                                                                                   &                                                                                  &                                                                                                    \\
                                     &                                                                                                   &                                                                                  &                                                                                                    \\
\multirow{3}{*}{AES-256}             & \multirow{3}{*}{\begin{tabular}[c]{@{}l@{}}3.5ns, WNS: 1.003ns\\ Setup failed\end{tabular}}       & \multirow{3}{*}{\begin{tabular}[c]{@{}l@{}}Pin placement \\ failed\end{tabular}} & \multirow{3}{*}{\begin{tabular}[c]{@{}l@{}}Stripes Not \\ Connected to \\ Power Ring\end{tabular}} \\
                                     &                                                                                                   &                                                                                  &                                                                                                    \\
                                     &                                                                                                   &                                                                                  &                                                                                                    \\
\multirow{3}{*}{NVDLA-small}         & \multirow{3}{*}{\begin{tabular}[c]{@{}l@{}}2ns, WNS: 0.65ns\\ Setup failed\end{tabular}}          & \multirow{3}{*}{\begin{tabular}[c]{@{}l@{}}Pin placement \\ failed\end{tabular}} & \multirow{3}{*}{\begin{tabular}[c]{@{}l@{}}Missing Bottom \\ and Top Power \\ Rings\end{tabular}}  \\
                                     &                                                                                                   &                                                                                  &                                                                                                    \\
                                     &                                                                                                   &                                                                                  &                                                                                                    \\
\multirow{3}{*}{NVDLA-large}         & \multirow{3}{*}{\begin{tabular}[c]{@{}l@{}}2.5ns, WNS: 0.69ns\\ Setup failed\end{tabular}}        & \multirow{3}{*}{\begin{tabular}[c]{@{}l@{}}Pin placement \\ failed\end{tabular}} & \multirow{3}{*}{\begin{tabular}[c]{@{}l@{}}Missing Bottom \\ and Top Power \\ Rings\end{tabular}}  \\
                                     &                                                                                                   &                                                                                  &                                                                                                    \\
                                     &                                                                                                   &                                                                                  &                                                                                                    \\
\multirow{3}{*}{UART controller}     & \multirow{3}{*}{\begin{tabular}[c]{@{}l@{}}3.4ns, WNS: 1.108ns\\ Setup successfully\end{tabular}} & \multirow{3}{*}{\begin{tabular}[c]{@{}l@{}}Util: 0.65\\ DRC Vio: 0\end{tabular}} & \multirow{3}{*}{\begin{tabular}[c]{@{}l@{}}Stripes Not \\ Connected to \\ Power Ring\end{tabular}} \\
                                     &                                                                                                   &                                                                                  &                                                                                                    \\
                                     &                                                                                                   &                                                                                  &                                                                                                    \\
\multirow{3}{*}{Elevator FSM}        & \multirow{3}{*}{\begin{tabular}[c]{@{}l@{}}2.5ns, WNS: 0.708ns\\ Setup successfully\end{tabular}} & \multirow{3}{*}{\begin{tabular}[c]{@{}l@{}}Util: 0.7\\ DRC Vio: 0\end{tabular}}  & \multirow{3}{*}{\begin{tabular}[c]{@{}l@{}}Stripes Not \\ Connected to \\ Power Ring\end{tabular}} \\
                                     &                                                                                                   &                                                                                  &                                                                                                    \\
                                     &                                                                                                   &                                                                                  &                                                                                                    \\
\multirow{3}{*}{Ethernet MAC}        & \multirow{3}{*}{\begin{tabular}[c]{@{}l@{}}10ns, WNS: 5.621ns\\ Setup failed\end{tabular}}        & \multirow{3}{*}{\begin{tabular}[c]{@{}l@{}}Pin placement \\ failed\end{tabular}} & \multirow{3}{*}{\begin{tabular}[c]{@{}l@{}}Missing Bottom \\ and Top Power \\ Rings\end{tabular}}  \\
                                     &                                                                                                   &                                                                                  &                                                                                                    \\
                                     &                                                                                                   &                                                                                  &                                                                                                    \\ \hline
\end{tabular} }

\end{table}